\documentclass[twocolumn,epjc3mod]{svjour3}  
\RequirePackage{flushend}
\smartqed  
\RequirePackage{graphicx}

\RequirePackage[numbers,sort&compress]{natbib}
\RequirePackage[colorlinks,citecolor=blue,urlcolor=blue,linkcolor=blue]{hyperref}

\RequirePackage[separate-uncertainty,bracket-numbers = false,
  table-align-uncertainty = true, table-align-exponent=false, exponent-product=\cdot]{siunitx}
\RequirePackage{dcolumn}
\RequirePackage{hyperref}
\RequirePackage{amssymb}
\RequirePackage{placeins}
\RequirePackage{subfigure}
\RequirePackage{amsmath}
\RequirePackage{amssymb}
\RequirePackage{upgreek}
\RequirePackage{multirow}
\RequirePackage[utf8]{inputenc}
\RequirePackage[british]{babel}
\RequirePackage{csquotes}
\RequirePackage{ulem}
\RequirePackage{cuted}
\RequirePackage{float}
\usepackage{xcolor}

\newcommand{\subfigureautorefname}{\figureautorefname}

\catcode`\@=11 
\def\parenbar{\mathpalette\p@renb@r}
\def\p@renb@r#1#2{\vbox{%
		\ifx#1\scriptscriptstyle \dimen@.7em\dimen@ii.2em\else
		\ifx#1\scriptstyle \dimen@.8em\dimen@ii.25em\else
		\dimen@1em\dimen@ii.4em\fi\fi \offinterlineskip
		\ialign{\hfill##\hfill\cr
			\vbox{\hrule width\dimen@ii}\cr
			\noalign{\vskip-.3ex}%
			\hbox to\dimen@{$\mathchar300\hfil\mathchar301$}\cr
			\noalign{\vskip-.3ex}%
			$#1#2$\cr}}}
\catcode`\@=12 
\def\nuan{\parenbar{\nu}\kern-0.4ex}

\newcommand{\overbar}[1]{\mkern 1.5mu\overline{\mkern-1.5mu#1\mkern-1.5mu}\mkern 1.5mu}
\newcommand{\anu}{\overbar{\nu}}
\newcommand{\nue}{\nu_{\text{e}}}
\newcommand{\numu}{\nu_{\mu}}

\newcommand{\nuane}{\protect\nuan_{e}}
\newcommand{\nuanmu}{\protect\nuan_{\mu}}
\newcommand{\nuantau}{\protect\nuan_{\tau}}
\newcommand{\anue}{\overbar{\nu}_{e}}
\newcommand{\anumu}{\overbar{\nu}_{\mu}}

\newcommand{\nueCC}{\nu_{\text{e}}\,\text{CC}}
\newcommand{\numuCC}{\nu_{\mu}\,\text{CC}}
\newcommand{\nutauCC}{\nu_{\tau}\,\text{CC}}
\newcommand{\nuNC}{\nu\,\text{NC}}

\newcommand{\anueCC}{\overbar{\nu}_{e}\,\text{CC}}
\newcommand{\anumuCC}{\overbar{\nu}_{\mu}\,\text{CC}}
\newcommand{\anuNC}{\overbar{\nu}\,\text{NC}}
\newcommand{\anutauCC}{\overbar{\nu}_{\tau}\,\text{CC}}
\newcommand{\nuaneCC}{\protect\nuan_{e}\,\text{CC}}
\newcommand{\nuanmuCC}{\protect\nuan_{\mu}\,\text{CC}}
\newcommand{\nuantauCC}{\protect\nuan_{\tau}\,\text{CC}}
\newcommand{\nuanNC}{\protect\nuan\ \text{NC}}
\newcommand{\nuanemuCC}{\protect\nuan_{\text{e},\mu}\,\text{CC}}
\newcommand{\nuemuCC}{\protect\nu_{\text{e},\mu}\,\text{CC}}
\newcommand{\anuemuCC}{\protect\overbar{\nu}_{\text{e},\mu}\,\text{CC}}
\newcommand{\dmsq}[1]{\Delta m^2_{#1}}
\newcommand{\deltaCP}{\delta_{\text{CP}}}
\newcommand{\sinsq}[1]{{\mathrm{sin}^2\theta_{#1}}}
\newcommand{\lik}{\ensuremath{LL}}

\newcommand{\pc}[1]{#1}

\let\orgautoref\autoref
\providecommand{\Autoref}{%
	\def\equationautorefname{Equation}%
	\def\figureautorefname{Figure}%
	\def\subfigureautorefname{Figure}%
	\def\subsectionautorefname{Section}%
	\def\sectionautorefname{Section}%
	\def\chapterautorefname{Section}%
	\def\tableautorefname{Table}
	\orgautoref}
\renewcommand{\autoref}{%
	\def\equationautorefname{Equation}%
	\def\figureautorefname{Figure}%
	\def\subfigureautorefname{Figure}%
	\def\subsubfigureautorefname{Figure}%
	\def\subsectionautorefname{Section}%
	\def\subsubsectionautorefname{Section}%
	\def\sectionautorefname{Section}
	\def\chapterautorefname{Section}
	\def\tableautorefname{Table}%
	\orgautoref}

\graphicspath{ {plots/}{plots/detector_performance/} }

\newcommand{\nmoNO}{\num{4.4}}

\newcommand{\nmoIO}{\num{2.3}}

\newcommand{\yNO}{\num{1.3}}

\newcommand{\yIO}{\num{5.0}}

\newcommand{\dThErrNO}{\ang[parse-numbers  = false ]{(^{+1.9}_{-3.1})}}

\newcommand{\dmErrNO}{\SI{85d-6}{eV^2}}

\newcommand{\dmErrIO}{\SI{75d-6}{eV^2}}

\newcommand{\dThErrIO}{\ang[parse-numbers  = false ]{(^{+2.0}_{-7.0})}}

\journalname{Eur. Phys. J. C}

\begin{document}\sloppy
\title{Determining the Neutrino Mass Ordering and Oscillation Parameters with KM3NeT/ORCA}


\author{
S.~Aiello\thanksref{a}
\and
A.~Albert\thanksref{bc,b}
\and
S. Alves Garre\thanksref{c}
\and
Z.~Aly\thanksref{d}
\and
A. Ambrosone\thanksref{e,f}
\and
F.~Ameli\thanksref{g}
\and
M.~Andre\thanksref{h}
\and
G.~Androulakis\thanksref{i}
\and
M.~Anghinolfi\thanksref{j}
\and
M.~Anguita\thanksref{k}
% ...
\and
(the KM3NeT Collaboration)}

\institute{\setlength{\parindent}{0em}
\label{a}INFN, Sezione di Catania, Via Santa Sofia 64, Catania, 95123 Italy
\and
\label{b}Universit{\'e}~de~Strasbourg,~CNRS,~IPHC~UMR~7178,~F-67000~Strasbourg,~France
\and
\label{c}IFIC - Instituto de F{\'\i}sica Corpuscular (CSIC - Universitat de Val{\`e}ncia), c/Catedr{\'a}tico Jos{\'e} Beltr{\'a}n, 2, 46980 Paterna, Valencia, Spain
\and
\label{d}Aix~Marseille~Univ,~CNRS/IN2P3,~CPPM,~Marseille,~France
\and
\label{e}INFN, Sezione di Napoli, Complesso Universitario di Monte S. Angelo, Via Cintia ed. G, Napoli, 80126 Italy
\and
\label{f}Universit{\`a} di Napoli ``Federico II'', Dip. Scienze Fisiche ``E. Pancini'', Complesso Universitario di Monte S. Angelo, Via Cintia ed. G, Napoli, 80126 Italy
\and
\label{g}INFN, Sezione di Roma, Piazzale Aldo Moro 2, Roma, 00185 Italy
\and
\label{h}Universitat Polit{\`e}cnica de Catalunya, Laboratori d'Aplicacions Bioac{\'u}stiques, Centre Tecnol{\`o}gic de Vilanova i la Geltr{\'u}, Avda. Rambla Exposici{\'o}, s/n, Vilanova i la Geltr{\'u}, 08800 Spain
\and
\label{i}NCSR Demokritos, Institute of Nuclear and Particle Physics, Ag. Paraskevi Attikis, Athens, 15310 Greece
\and
\label{j}INFN, Sezione di Genova, Via Dodecaneso 33, Genova, 16146 Italy
\and
\label{k}University of Granada, Dept.~of Computer Architecture and Technology/CITIC, 18071 Granada, Spain
% ...
\and
\label{bc}Universit{\'e} de Strasbourg, Universit{\'e} de Haute Alsace, GRPHE, 34, Rue du Grillenbreit, Colmar, 68008 France
% ...
}

\thankstext{corr1}{e-mail: steffen.hallmann@fau.de}
\thankstext{corr2}{e-mail: jannik.hofestaedt@fau.de}
\thankstext{corr3}{e-mail: mathieu.perrin-terrin@cppm.in2p3.fr}

\date{}

\maketitle

\begin{abstract}
\pc{The next generation of water Cherenkov neutrino telescopes in the Mediterranean Sea are under construction offshore France (KM3NeT/ORCA) and Sicily (KM3NeT/ARCA). The KM3NeT/ORCA detector features an energy detection threshold which allows to collect atmospheric neutrinos to study flavour oscillation. This paper reports the KM3NeT/ORCA sensitivity to this phenomenon.} The event reconstruction, selection and
classification are described.
The sensitivity to determine the neutrino mass ordering was evaluated and found to be \nmoNO $\sigma$ if the true ordering is normal and \nmoIO $\sigma$ if inverted, after three years of data taking. The precision to measure $\dmsq{32}$ and $\theta_{23}$ were also estimated and found to be \dmErrNO\ and \dThErrNO\ for normal neutrino mass ordering and, \dmErrIO\ and \dThErrIO\ for inverted ordering.
Finally, a unitarity test of the leptonic mixing matrix by measuring the rate of tau neutrinos is described. Three years of data taking were found to be sufficient to exclude $\nuantau$ event rate variations larger than 20\% at $3\sigma$ level.
\end{abstract}

\section{Introduction}

The standard framework of three neutrino flavour eigenstates ($\nu_e$, $\nu_\mu$, $\nu_\tau$), which are superpositions of the three mass eigenstates ($\nu_1$, $\nu_2$, $\nu_3$) with masses ($m_1$, $m_2$, $m_3$), has been established with more than two decades of neutrino oscillation physics research.
By convention, $\nu_1$ is the mass eigenstate with the largest $\nu_e$ component, and $\nu_3$ is the one with the smallest. 
The ordering of the neutrino mass eigenstates is not yet resolved,
and it can be either 
$m_1<m_2<m_3$ (`normal ordering', NO) 
or $m_3<m_1<m_2$ (`inverted ordering', IO).
The question of the neutrino mass ordering (NMO) is one of the main drivers of neutrino oscillation physics. 

Neutrino mixing is described by the Pontecorvo-Maki-Nakagawa-Sakata (PMNS) matrix, $U$, \cite{MakiEtAl_1962,Pontecorvo_1968,GribovEtAl_1969} with
\begin{equation}
 \nu_\alpha  = \sum_{i=1}^3 U_{\alpha i}\nu_i,
\end{equation} 
where $\alpha = {e,\mu,\tau}$ and $i = {1,2,3}$.
Under the assumption that the mixing matrix $U$ is unitary, it is usually parametrised in terms of three mixing angles $\theta_{12}$, $\theta_{13}$ and $\theta_{23}$, and a CP-violating phase $\deltaCP$ \cite{PDG2018}.
Neutrino oscillations are sensitive to mass-squared differences $\Delta m_{ij}^2 = m^2_i - m^2_j ~ (i,j=1,2,3)$.
From the three neutrino mass eigenstates two independent mass-squared differences can be constructed, 
which we choose as $\Delta m_{12}^2$ and $\pm \left |\Delta m_{23}^2 \right|$, where the sign of the latter is positive for NO and negative for IO. 

Global fits of the available data form a coherent picture and provide values for $\theta_{12}$, $\theta_{13}$, $\theta_{23}$, $\Delta m_{12}^2$ and $\left |\Delta m_{23}^2 \right|$ with few-percent level precision \cite{globalFitSalas,globalFitEsteban,globalFitCapozzi}.
However, some questions remain:
the determination of
the value of $\delta_{\rm{CP}}$,
the octant of $\theta_{23}$ (i.e. whether $\theta_{23}$ is greater or smaller than $\pi/4$) 
and the neutrino mass ordering (i.e. the sign of $\Delta m_{23}^2$).
The current status is that global fits \cite{Esteban:2020cvm,Kelly:2020fkv} indicate a mild preference for 
NO over IO, second octant of $\theta_{23}$ and $\delta_{\rm{CP}} \approx \pi$ to $\frac{3}{2} \pi$.
The experiments driving the NMO sensitivity results are T2K \cite{hep-ph_T2K_AbeEtAl_2020}, NOvA \cite{hep-ph_NOvA_AceroEtAl_2019}, MINOS \cite{aurisano_adam_2018_1286760}, Super-Kamiokande \cite{hep-ph_SuperKamiokande_2018} and IceCube/DeepCore \cite{DeepCore2017}. 
Notably, the hints for NO tend to weaken in the light of combined analyses \cite{Esteban:2020cvm,Kelly:2020fkv} 
using the latest results from T2K \cite{neutrino2020talk_T2K} and NOvA \cite{neutrino2020talk_NOvA}.

Deriving strong experimental constraints on the unitarity of the $3\times3$ PMNS mixing matrix is challenging, as direct observations of $\nuanmu\rightarrow\nuantau$  are difficult and the $\tau$ rest-mass suppresses the $\nuantau$ interaction cross section.
Appearance of $\nuantau$ has been directly observed at the long baseline CNGS neutrino beam by OPERA \cite{tau_appearance_opera_discovery2015,tau_appearance_opera_finalresults2018}. Evidence for $\nuantau$ appearance has also been found on a statistical basis in the atmospheric neutrino flux by Super-Kamiokande \cite{tau_appearance_superk_2017} and IceCube \cite{tau_appearance_icecube_2019}. 
However, the uncertainty on the normalisation of the $\nuantau$ signal is currently too large to probe the unitarity of the PMNS mixing matrix.\pc{ Non-unitarity would imply the incompleteness of the $3\times3$ flavour paradigm and could point to the existence of additional neutrino flavours.} A statistically highly-significant detection of $\nuantau$ appearance from $\nuanmu \rightarrow \nuantau$ oscillations of atmospheric neutrinos could make an important contribution to further constrain the PMNS matrix elements involving $\nuantau$. 

The NMO can be determined by measuring the energy and zenith angle dependent oscillation pattern of few-GeV atmospheric neutrinos that have traversed the Earth \cite{Akhmedov2013}. 
Matter-induced modifications \cite{Wolfenstein1978,Mikheev:1986gs} of the oscillation probabilities lead to an enhancement of the 
$\nuanmu \leftrightarrow \nuane$
transition for neutrinos in the case of NO, and anti-neutrinos in the case of IO.
\pc{Earth matter effects are due to coherent neutrino electron forward scattering. They arise mainly below $E_\nu \lesssim 15$\,GeV and depend on the electron density of the medium. The largest effects appear around $7$\,GeV for neutrinos passing through the Earth’s mantle and around $3$\,GeV for neutrinos passing through the Earth’s core.}
The oscillation pattern for neutrinos with respect to anti-neutrinos is flipped between the two mass orderings.

In case of detectors that cannot distinguish between neutrinos and anti-neutrinos on an event-by-event basis, the determination of the NMO can be based on the observation of a net difference in the event rates of atmospheric neutrinos, resulting from a higher interaction cross section (factor $\sim 2$) and the existing atmospheric flux difference (factor $\sim 1.1$) for neutrinos with respect to anti-neutrinos.
Due to this event rate difference, the strength of the observed matter effects,
i.e. the enhancement of the $\nuanmu \leftrightarrow \nuane$ transition,
is larger for NO compared to IO.
This is the experimental signature exploited by KM3NeT/ORCA and other atmospheric neutrino experiments to determine the NMO.

KM3NeT is a large research infrastructure that will consist of a network of deep-sea neutrino detectors in the Mediterranean Sea.
Two underwater neutrino telescopes, called ARCA and ORCA, are currently under construction \cite{LoI}.
ARCA (Astroparticle Research with Cosmics in the Abyss) is a sparsely instrumented gigaton-scale detector optimised for TeV--PeV neutrino astronomy.
ORCA (Oscillation Research with Cosmics in the Abyss) is a more densely instrumented detector 
optimised for measuring the oscillation of few-GeV atmospheric neutrinos in order to determine the neutrino mass ordering.

{
With atmospheric neutrino data, ORCA can also perform a precise measurement of $\theta_{23}$ and $\Delta m_{23}^2$ as well as a high-statistics measurement of $\nuantau$ appearance in the atmospheric neutrino flux,
which allows to probe deviations from the unitarity assumption of the $3$-neutrino mixing.
Sensitivity for tau-neutrino appearance mainly comes from atmospheric neutrinos with energy $\gtrsim\SI{15}{GeV}$
and therefore has only a weak dependence on the still undetermined neutrino mass ordering.}

A first estimation of the sensitivity of ORCA to the NMO as well as to other oscillation parameters was published in the `Letter of Intent for KM3NeT 2.0' (LoI) \cite{LoI}. 
\pc{Since then, the detector and the analysis methods have been further optimised. First, the detector geometry has been updated. In addition, significant improvements in the neutrino detection efficiency as well as reconstruction performance have been achieved as illustrated in \autoref{sec:reco}. The event classification procedure has been significantly improved as well. We use now three event classes and hit features are included, this is discussed in \autoref{sec:classification}. At the same time the analysis has been refined. The detector response is modeled in greater detail and a more complete list of systematic effects is now considered. These effects partly compensate the expected gain in sensitivity from the improvements mentioned above but make them at the same time more realistic.}
The updated sensitivities are presented in this paper.

This paper is organised as follows. 
\Autoref{Sec:detector} describes the detector design and the simulations performed to obtain the detector response to atmospheric neutrinos, atmospheric muons as well as optical background noise. Then, the algorithms used for event reconstruction and for high flavour purity event classification are described.
In \autoref{Sec:analysis}, the methods used to analyse these samples and derive the sensitivity to the NMO, the atmospheric oscillation parameters and the $\nuantau$ appearance are presented together with the results.
Finally, \autoref{Sec:conclusion} summarises the main detector and analysis updates and the expected sensitivity to neutrino oscillations.

\label{Sec:Introduction}

\section{ORCA Detector Response}
\label{Sec:detector}
The ORCA detector design comprises a 3-dimensional array of photosensors 
that register the Cherenkov light produced by relativistic charged particles 
emerging from neutrino-induced interactions.  
The arrival time of the Cherenkov photons
and the position of the sensors
are used to reconstruct
the energy and direction of the incoming neutrino
as well as the event topology.

\subsection{Detector Design}
The ORCA detector design consists of an array of 115 vertical detection units (DUs) 
featuring 18 digital optical modules (DOMs) each.
Each DOM is a pressure-resistant glass sphere, housing 31 photomultiplier tubes (PMTs) of 3-inch diameter and the related electronics.
The KM3NeT PMTs are characterised in \cite{Aiello2018}.

The detector is located at the KM3NeT-France site and the base container of each DU is placed at about $\SI{2450}{m}$ depth.
The DUs are arranged in a circular footprint with a radius of about $\SI{115}{m}$
with an average spacing between the DUs of $\SI{20}{m}$. 
Along a DU, the vertical spacing between the DOMs 
varies between $\SI{8.7}{m}$ to $\SI{10.9}{m}$ (due to technical constraints from the deployment procedure)
with an average of $\SI{9.3}{m}$. 
The first DOM is at a distance of about $\SI{30}{m}$ from the seabed \cite{Aiello2020KM3NeTdeployment}.
In total, a volume of about $\SI{6.7e6}{m^3}$ (equivalent to $\SI{7.0}{Mt}$ of sea water) is instrumented.
This detector configuration is the outcome of an optimisation study using the sensitivity to the NMO as figure of merit. 

\subsection{Simulation} \label{Sec:Simulation}
Detailed Monte Carlo (MC) simulations are used to evaluate the detector response to atmospheric neutrinos, atmospheric muons and optical background noise. 
The simulation chain used for the analysis presented in this paper is similar to the one described in \cite{LoI}. 

Neutrino induced interactions in sea water are simulated with gSeaGen \cite{gSeaGen_paper}, a software package based on the widely used GENIE (version 2.12.10) code \cite{GENIE,GENIE_manual}.
Neutrinos and antineutrinos in the energy range from $\SI{1}{GeV}$ to $\SI{100}{GeV}$ are simulated and weighted to reproduce the conventional atmospheric
neutrino flux following the Honda model \cite{Honda2015}.
All particles emerging from neutrino interactions are propagated with the GEANT4-based software package KM3Sim \cite{KM3Sim}. 
Using this software, Cherenkov photons are generated from primary and secondary particles, tracked through the sea water taking into account absorption and scattering, and detected by the PMTs. 

Atmospheric muon events are generated using the MUPAGE package \cite{MUPAGE}.
The KM3 package \cite{Bailey:2002uj,AntaresSimulation2020} is then used for tracking the muons in sea water and the subsequent Cherenkov light production.

The PMT response and the readout are simulated using custom KM3NeT software. 
The digitised PMT output signal is typically called a {\it hit}. 
In this step, the optical background due to Cherenkov light from $\beta$-decays of $^{40}$K in the sea water is also added:
an uncorrelated hit rate of $\SI{10}{kHz}$ per PMT as well as time-correlated noise on multiple PMTs on each DOM ($\SI{600}{Hz}$ twofold, $\SI{60}{Hz}$ threefold, $\SI{7}{Hz}$ fourfold, $\SI{0.8}{Hz}$ fivefold and $\SI{0.08}{Hz}$ sixfold). 
The simulated time-correlated noise rate is taken from the data of the first deployed DUs
\cite{depthIntensityRelationPaper}.
Finally, the simulated data is filtered by dedicated trigger algorithms to identify events induced by energetic particles. 
The trigger algorithms are designed to search for large clusters of causally-connected hits.
The same trigger algorithms are applied to both simulated and real data.

Compared to the LoI \cite{LoI},
significant improvements have been made in the triggering of faint events with only a few tens of detected photons \cite{SteffenThesis}.  
A new trigger algorithm has been developed for the needs of ORCA.
It is based on only one local coincidence 
(photons recorded on two or more PMTs of the same DOM 
within 10\,ns) 
and a tunable number of causally-connected single hits on DOMs in the vicinity. A minimum of seven additional hits distributed over at least three different DOMs are required.
This new algorithm significantly increases the trigger efficiency in the few-GeV neutrino energy range, while still satisfying the bandwidth requirements of the data acquisition system. 

The total trigger rate due to atmospheric muons is about $\SI{50}{Hz}$ and noise events add about $\SI{54}{Hz}$,
while atmospheric neutrinos are triggered with a rate of about $\SI{8}{mHz}$.
In total,
1.4 days of noise events, 14 days of atmospheric muons 
and more than 15 years of atmospheric neutrinos are simulated.
These event samples are sufficient to probe a percent-level background contamination (see \autoref{sec:classification}).
In future analysis of real data, the background will be included based on run-by-run simulations \cite{AntaresSimulation2020},
accounting for the detector and data-taking conditions. 

\subsection{Event Topologies}

Two distinct event topologies can be distinguished in the detector: 
{\it track-like} and {\it shower-like}.
In the few-GeV energy range,
muons are the only particles that can be confidently identified, because they
are the only particles that appear as {\it tracks} in the detector, with a track length proportional to the muon
energy ($\sim$4\,m/GeV).
Electrons and hadrons initiate particle {\it showers} that develop over distances of a few metres. 
Compared to elongated muon tracks, these showers appear as localised light sources in the detector.
All neutrino-induced events producing a muon with sufficient energy are called
track-like, i.e. $\nuan_\mu$ charged-current (CC) events and $\nuantauCC$ events with muonic $\tau$ decays. All other neutrino-induced events are called shower-like, i.e. $\nuan_{e,\mu,\tau}$ neutral-current (NC) events, $\nuaneCC$ events and $\nuantauCC$ events
with non-muonic $\tau$ decays.

\subsection{Event Reconstruction and Event Selection}
\label{sec:reco}
Dedicated reconstruction algorithms are applied for track-like and shower-like events as well as an event topology classification algorithm.
The track and shower reconstruction algorithms are described in \cite{LiamThesis} and \cite{JannikThesis}, respectively. 
Both reconstruction algorithms are maximum likelihood fits and reconstruct the energy and direction as well as interaction vertex position and time.
\pc{Events reconstructed as upgoing, i.e. with a negative cosine zenith angle, are selected based on the reconstruction quality and containment.} The containment criteria are based on the event position and direction inside the instrumented detector volume \cite{SteffenThesis}. 
The goal of the event preselection is to fulfil two main purposes:
suppress background events and select well-reconstructed events with a good reconstruction accuracy. 

\begin{figure}[htb]
    \centering
    \includegraphics[width=\columnwidth]{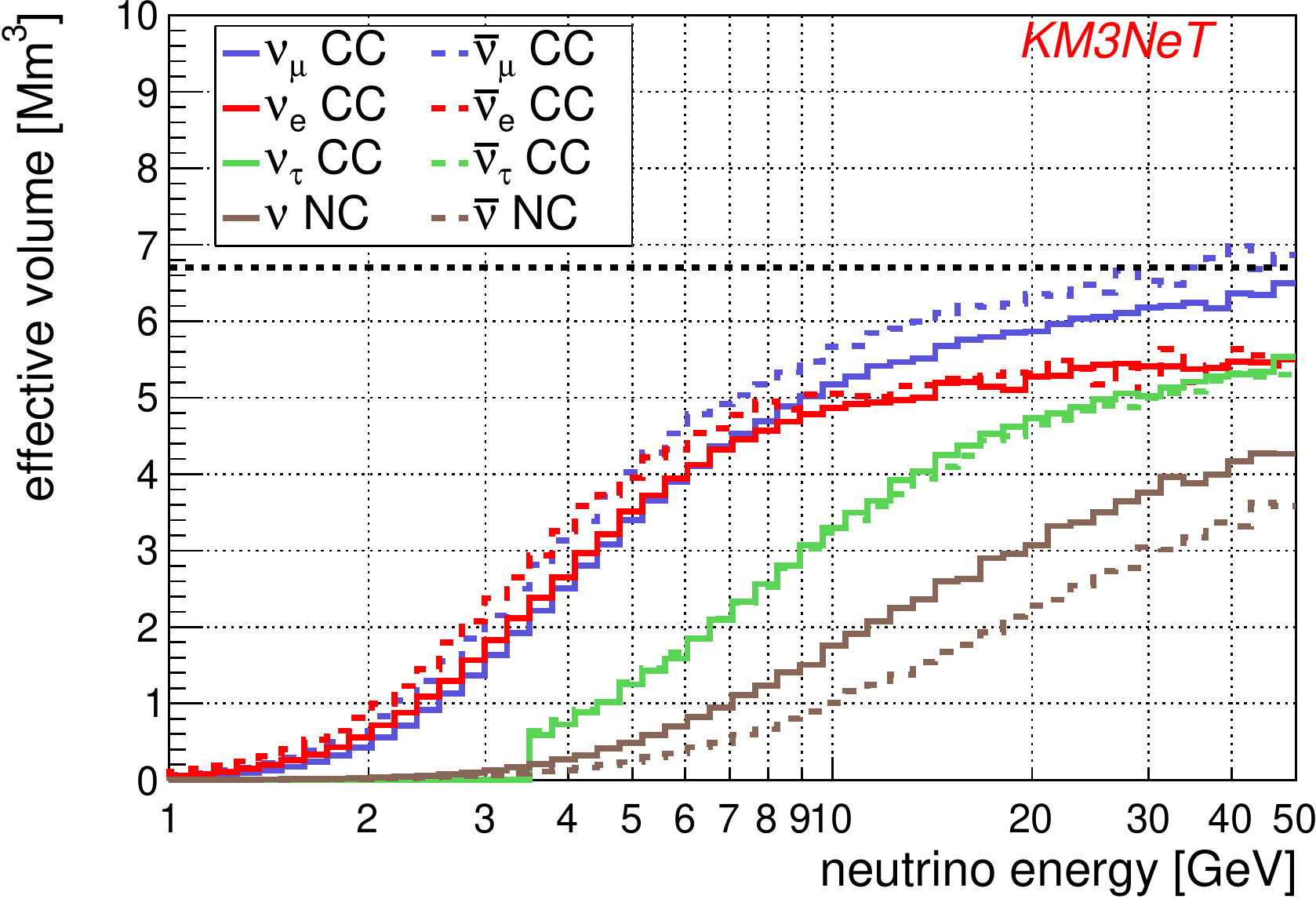}
    \caption{Effective detector volume as a function of true neutrino energy $E_\nu$ for different neutrino flavours and interactions.
    Events are weighted according to the Honda atmospheric neutrino flux model and averaged over the zenith angle.
    Only events reconstructed and selected as upgoing are used.
    The dashed black line indicates the instrumented volume of the detector.
    }
    \label{fig:effMass_nuTypes}
\end{figure}

The effective detector volume after the event preselection
is shown in \autoref{fig:effMass_nuTypes}
for upgoing neutrinos weighted according to the Honda atmospheric neutrino flux model \cite{Honda2015}.
The effective detector volume reaches a plateau and is nearly as large as the instrumented detector volume for $\nuanemuCC$ with $E_\nu \gtrsim \SI{15}{GeV}$, while 50\% efficiency is reached for $E_\nu \sim \SI{4}{GeV}$.
\pc{Compared  to  the LoI  \cite{LoI},  the  turn-on  region  of  the  effective  detector  volume  is  shifted  by  about  20\%  to  lower energies  due  to  improvements  in  event  triggering and  reconstruction. Indeed, as discussed in \autoref{Sec:Simulation}, additional methods have been developed to record events with a lower number of in-time hits from the same DOM but with extra hits causally connected on other DOMs and a similar method is applied at the prefit stage of the reconstruction. These refinements contribute to lower the detection energy threshold.}
In general, the effective volume is smaller for $\nuanNC$ and $\nuantauCC$ than for $\nuanemuCC$ events as the outgoing neutrinos are invisible to the detector. For $\anuemuCC$ events the effective volume is larger than for $\nuemuCC$ due to the lower average inelasticity and the resulting higher average light yield (at the considered energies hadronic showers
have a smaller average light yield than electromagnetic showers). The difference between $\nutauCC$ and $\anutauCC$ is diluted due to the effect of finite mass of the $\tau$ lepton on the neutrino interaction cross sections \cite{nutau_anutau_xsection}.
Due to the KM3NeT DOM design, more PMTs are oriented downwards (housed in the lower hemisphere) compared to oriented upwards (housed in the upper hemisphere), resulting in a higher photon detection efficiency for upgoing compared to horizontal events.

In total, a sample of about 66,000 upgoing neutrinos per year, 
corresponding to a rate of about 2\,mHz,
will be detected and can be used for further analysis.
In addition, about 0.4\,Hz of noise events 
and 0.1\,Hz of atmospheric muon events 
pass the preselection criteria.
To suppress the noise and atmospheric muon background,
a more sophisticated event classification is performed, 
as detailed in \autoref{sec:classification}.

\begin{figure*}[htb]
	\centering
	\includegraphics[width=.47\textwidth]{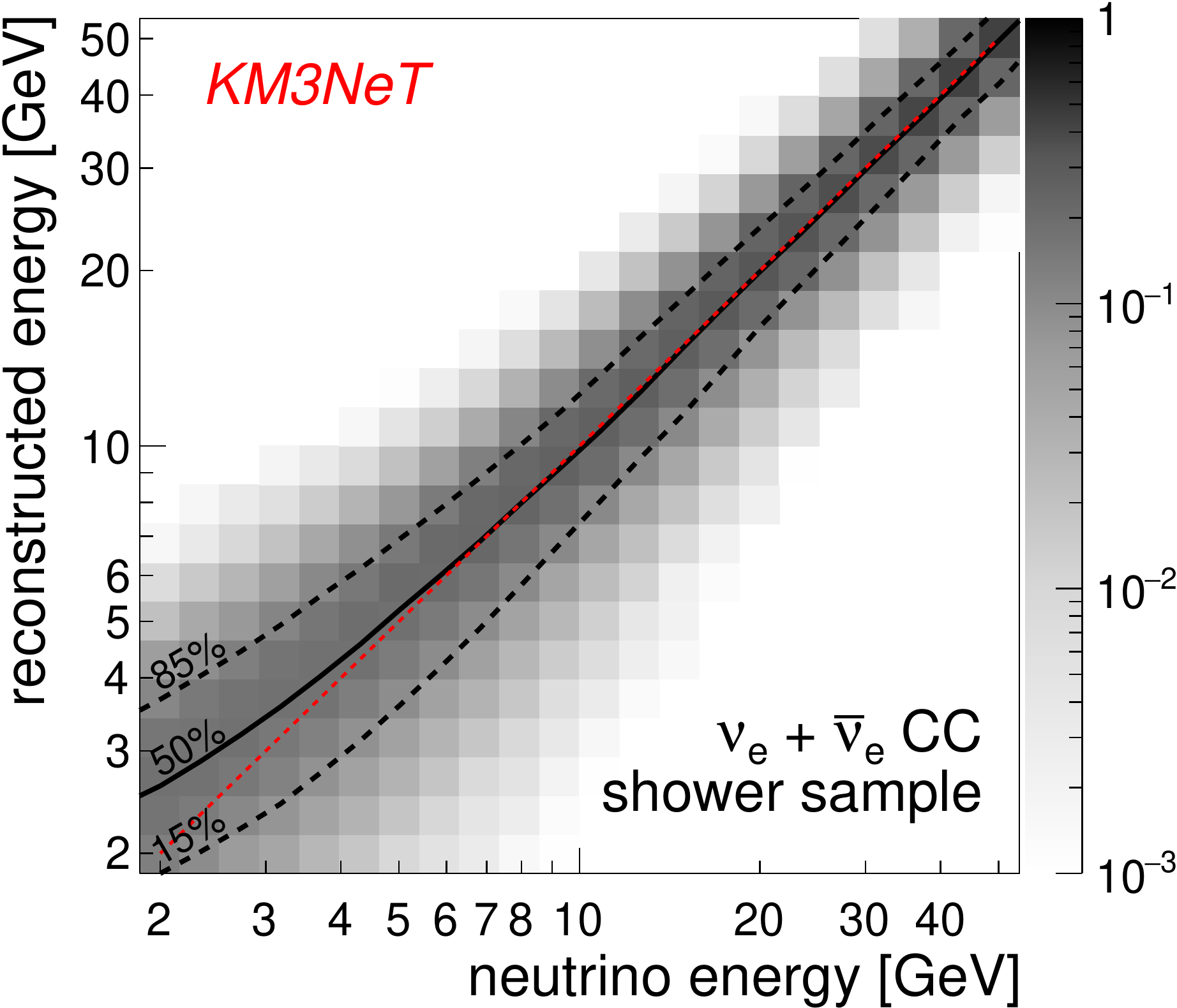}
	\hfill
	\includegraphics[width=.47\textwidth]{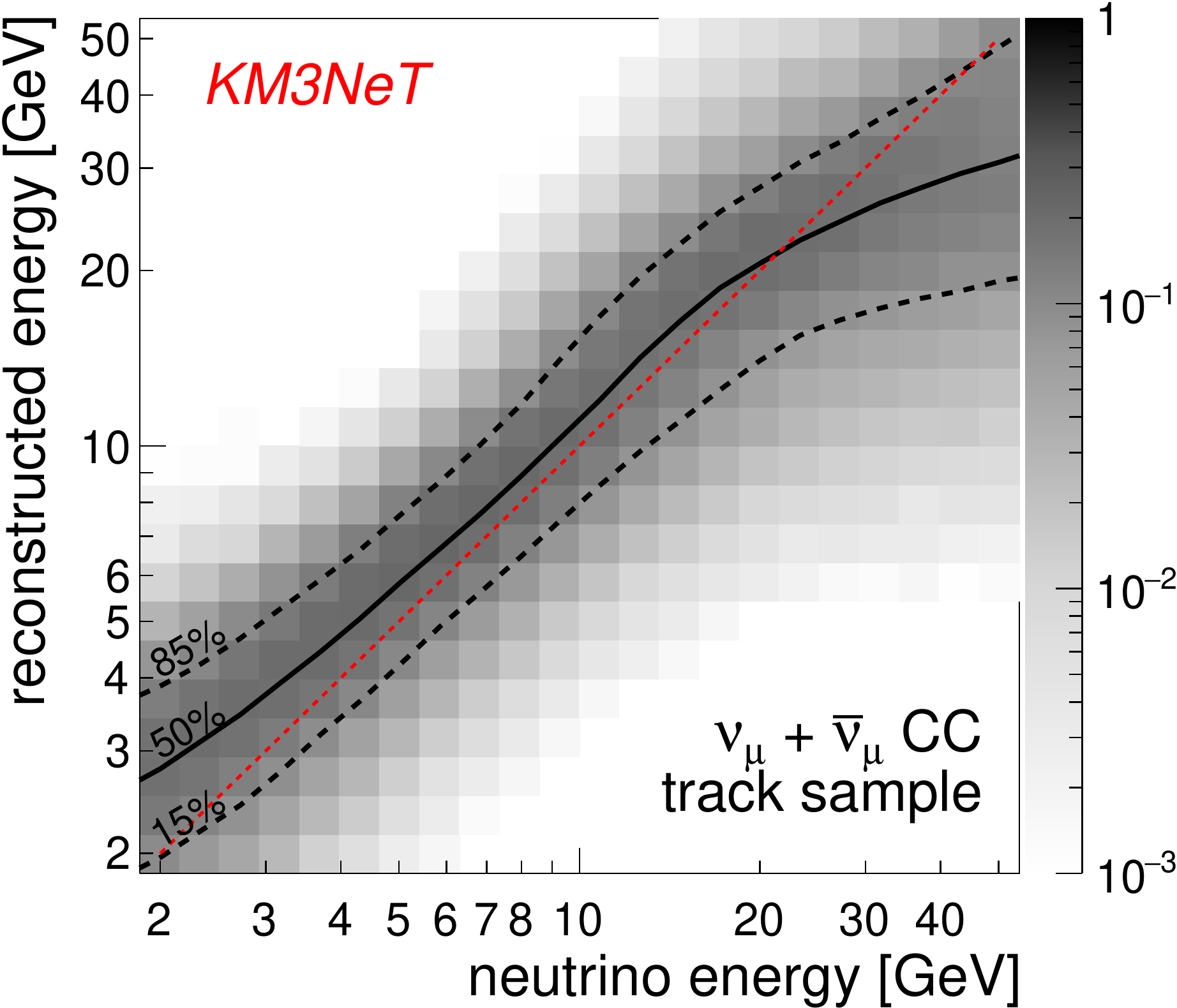}
	\caption{
	Probability distribution of the reconstructed energy as a function of true neutrino energy for upgoing $\nueCC$ and $\anueCC$ events classified as shower-like (left) as well as $\numuCC$ and $\anumuCC$ events classified as track-like (right). 
	Solid and dashed black lines indicate 50\%, 15\% and 85\% quantiles. For a definition of shower- and track-like events see \autoref{eq:classification}. 
    The red diagonal line indicates perfect energy reconstruction.
    } 
	\label{fig:resolution_E}
\end{figure*}

The energy resolution for $\nueCC$ and $\anueCC$ events classified as shower-like, as well as
$\numuCC$ and $\anumuCC$ events classified as track-like are shown in \autoref{fig:resolution_E}.
The energy resolution is Gaussian-like with $\Delta E / E \approx 25$\% for $\nuaneCC$ events with $E_\nu = \SI{10}{GeV}$,
and it is dominated by the intrinsic light yield fluctuations in the hadronic shower \cite{IntrinsicPaper}.
For $\nuanmuCC$, 
the resolution on the neutrino energy levels off at $\Delta E / E \approx 35$\% as the reconstructed muon track tends not to be fully contained inside the instrumented volume.

\begin{figure}[htb]
	\centering
	\includegraphics[width=\columnwidth]{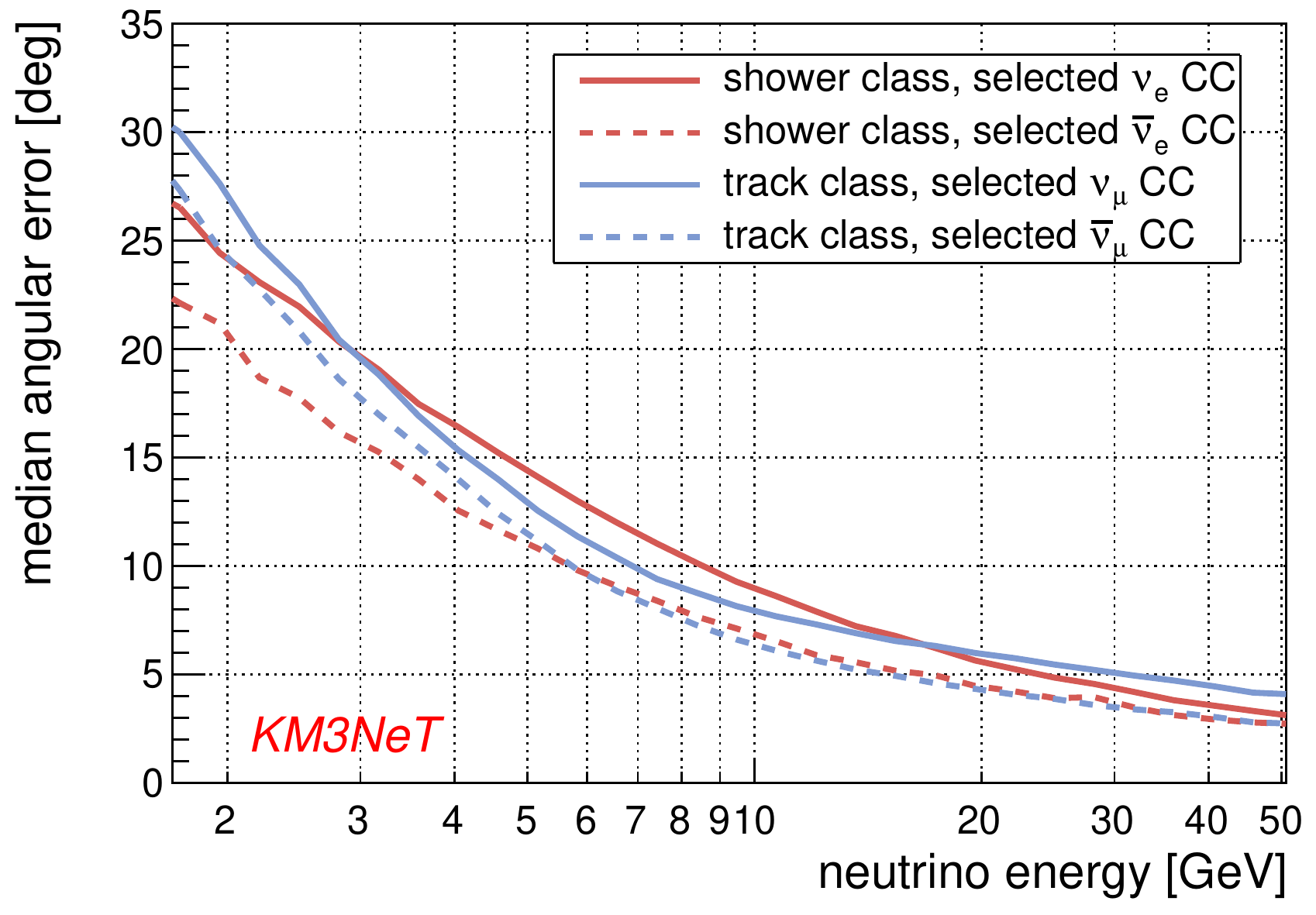}
	\caption{Median direction resolution as a function of true neutrino energy $E_\nu$ for upgoing $\nueCC$ and $\anueCC$ events classified as shower-like as well as $\numuCC$ and $\anumuCC$ events classified as track-like. For a definition of shower- and track-like events see \autoref{eq:classification}.} 
	\label{fig:resolution_dir}
\end{figure}

\autoref{fig:resolution_dir} shows the median resolution on the neutrino direction for the same set of simulated neutrino events.
At $E_\nu = \SI{10}{GeV}$, the median neutrino direction resolution is $9.3^\circ$/$7.0^\circ$/$8.3^\circ$/$6.5^\circ$ for $\nue$/$\anue$/$\numu$/$\anumuCC$ events, respectively. 
The neutrino direction resolution is dominated by the intrinsic $\nu$--lepton scattering kinematics \cite{IntrinsicPaper}, resulting in better resolutions for $\anu$ CC than for $\nu$ CC due to the smaller Bjorken-y.

\subsection{Event Classification}
\label{sec:classification}

For event classification, random decision forests (RDFs) \cite{Breiman2001} are used, which consist of an ensemble of binary decision trees.

Two RDFs are trained individually for selecting neutrino candidates against each of the two dominant classes of background -- atmospheric muons and noise events -- and a third one is trained to distinguish track-like from shower-like event topologies.

To train the classifiers, $\nuanmuCC$ events have been used to represent track-like event topologies. For showers $\nuaneCC$ and $\nuanNC$ events have been used. The neutrino event distributions were flattened in $\log_{10}$ of neutrino energy and the numbers of events per class were balanced between tracks and showers. 
In contrast, background was fed with the expected true spectra.

Each trained classifier yields a score variable (\texttt{atmospheric\_muon\_score}, \texttt{noise\_score}, \texttt{track\_score}). These represent the fraction of trees voting for the respective result class. The individual score parameters allow to separately optimise the suppression of the atmospheric muon and noise components using selection cuts and to divide the remaining events into different classes for analysis.

In the training, only events which pass the preselection requirements for either tracks or showers were used. The classifiers were trained independently of each other. Consequently, no further selection based on the resulting score from one of the other classifiers and none of the resulting score variables is used to train the RDFs.  
In the training, a forest size of 101 trees\footnote{\pc{The uneven number was chosen for practical purposes only and simplifies consistency of event selection across different analyses ($>$ vs. $\geq$)}}, and 50,000 events per class  (25,000 for noise suppression due to smaller available statistics after preselection) have been used. \pc{In the training process,  five-fold cross validation was applied.}

To ensure diversity of trees within the forest, each tree was trained on a randomly drawn 60\% subset of the training variables and 40\% of the available training events. 

The training variables consist of the fitted event parameters and additional variables quantifying the reconstruction quality. These are provided by the track and shower algorithms \cite{LiamThesis,JannikThesis}. Additional sets of variables fed to the classifier are relative distances between the fitted track and shower hypothesis and variables quantifying how well the Cherenkov light signature is contained within the instrumented volume. 

To separate between track- and shower-like signatures, further hit-based variables are added, which have not been used in \cite{LoI} and exploit the distribution of detected photon hits in the detector. These are based on likelihood ratios of the time and position of the hits expected for the $\nuaneCC$ and $\nuanmuCC$ event hypotheses with respect to the reconstructed position and direction of the shower reconstruction algorithm. More information on the classifier training can be found in \cite{SteffenThesis}.

\begin{figure*}[htb]
    \includegraphics[width=.47\textwidth]{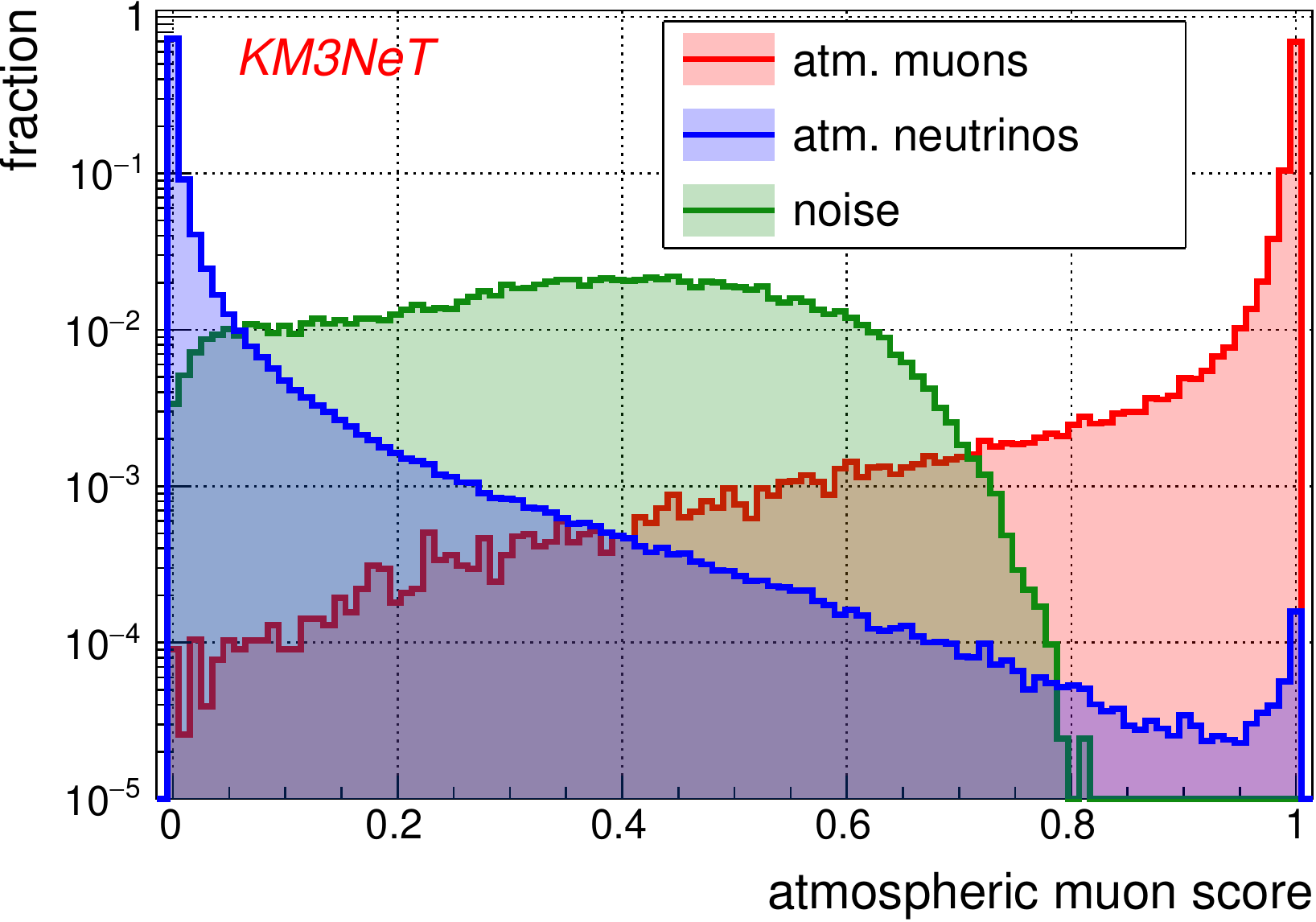}
    \hfill
    \includegraphics[width=.47\textwidth]{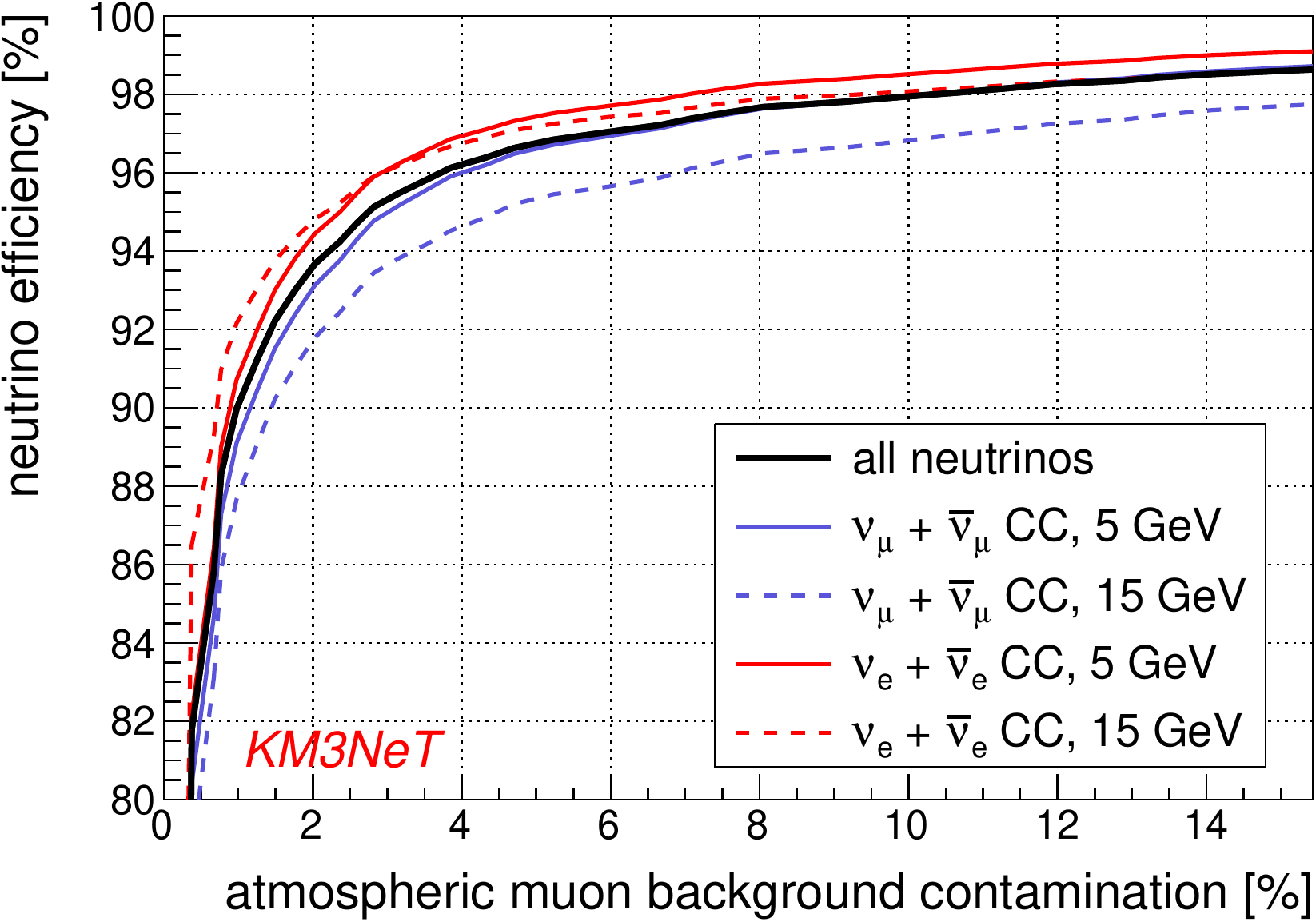}
    \caption{Left: Distribution of the atmospheric muon score variable for
        the RDF trained to separate between neutrinos and atmospheric muons,  for the main classes of events. Right: Fraction of remaining neutrinos weighted with an oscillated atmospheric flux versus atmospheric muon contamination in the final sample.
        }
    \label{fig:muon_suppression}
\end{figure*}
\begin{figure*}[htb]
    \includegraphics[width=.47\textwidth]{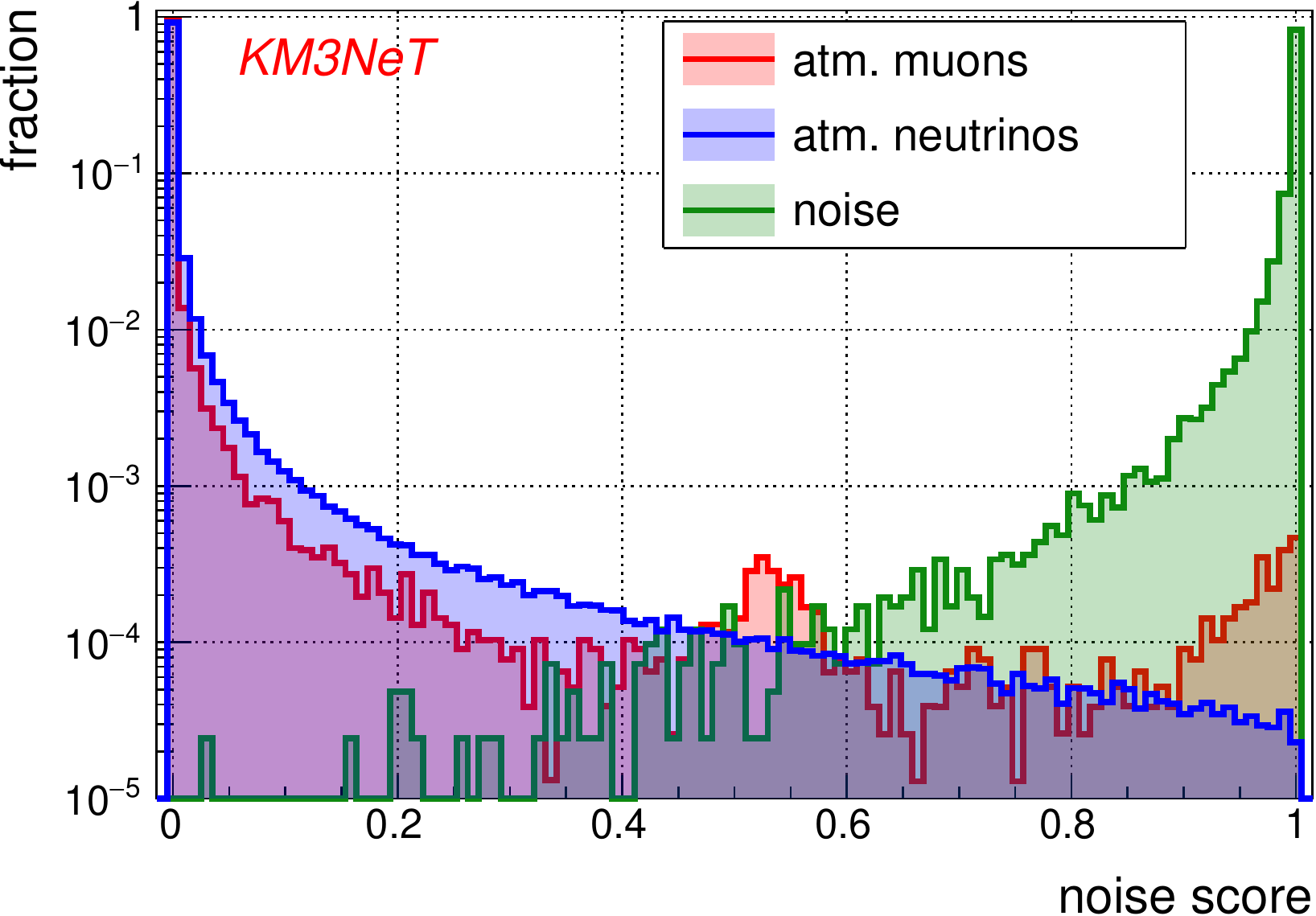}
    \hfill
    \includegraphics[width=.47\textwidth]{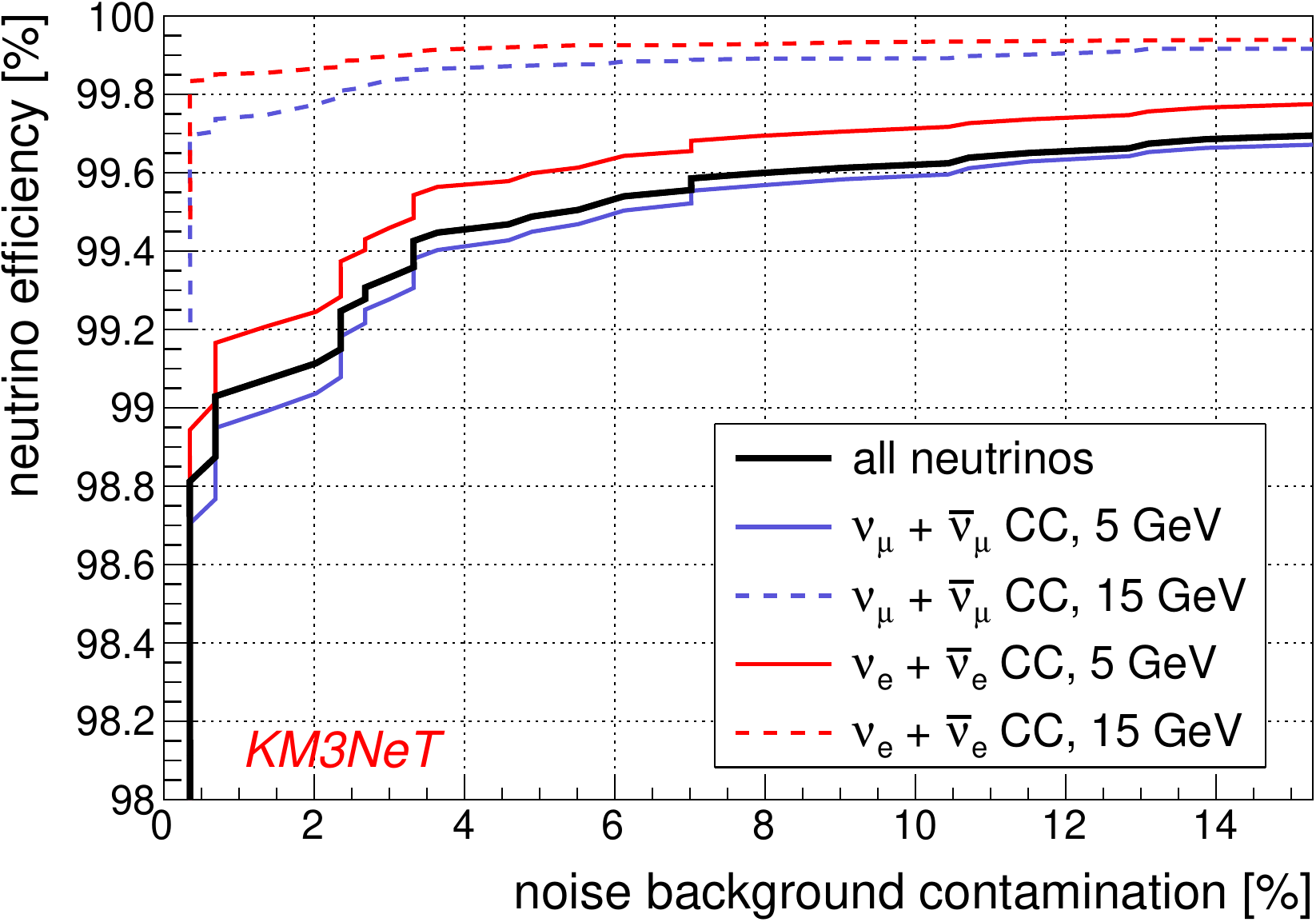}
    \caption{Left: Distribution of the noise score variable  for
        the RDF aimed to separate between neutrinos and pure noise, for the main classes of events. Right: Fraction of remaining atmospheric neutrinos versus noise event contamination in the final sample.
         }
    \label{fig:noise_suppression}
\end{figure*}

The classifier performance in rejecting the atmospheric muon background is given in \autoref{fig:muon_suppression}. The distribution of the \texttt{atmospheric\_muon\_score} (left panel) shows a clear separation between neutrinos weighted with an oscillated atmospheric flux and atmospheric muons. \pc{The increase of neutrino events with a $track_score \approx 1$ comes from $\nuanmu$ CC and $\nuantau$ CC events with $\tau^\pm$ decay to $\mu^\pm$ and is absent for other neutrino channels.} Noise events have not been used in training the classifier and therefore are not clustered at the edges of the distributions. 
A relatively hard cut at \texttt{atmospheric\_muon\_score} $< 0.05$ is used to reach a  $\sim3\%$ contamination level, cf. \autoref{fig:muon_suppression} (right panel). The loss in neutrino efficiency for the atmospheric muon rejection does not strongly depend on the neutrino energy and is about $\sim5\%$.

Noise events are rejected sufficiently with a cut on \texttt{noise\_score} $< 0.1$. As can be seen from \autoref{fig:noise_suppression} (right panel), the rejection of  noise events does not significantly reduce the number of neutrino events in the analysis sample. However, the reduction of neutrino events tends to increase
for faint neutrino events with energies near the detection threshold.
The proposed cuts on the \texttt{atmospheric\_muon\_score} and \texttt{noise\_score} values reduce the muon and noise contamination of the selected event sample to a level which can be safely neglected in the sensitivity study.

The training of track- versus shower-like neutrino event signatures results in a $\texttt{track\_score}$ variable, representing the fraction of trees voting for the candidate event to be track-like.
Using this variable, events can be split in three event classes based on the following criteria:
\begin{align}\label{eq:classification}
\text{shower\ class:}\quad &\text{passes\ shower\ preselection}\nonumber\\
        \quad \mathbf{and}\ &(\texttt{track\_score} \leq 0.3),\nonumber\\ 
\text{intermediate\ class:}\quad &\text{passes\ shower\ preselection}\nonumber\\
        \quad \mathbf{and}\ &(0.3 < \texttt{track\_score} \leq 0.7),\nonumber \\
\text{track\ class:}\quad &\text{passes\ track\ preselection}\nonumber\\
        \quad \mathbf{and}\ &(\texttt{track\_score} > 0.7).
\end{align}

\begin{figure*}[htb]
    \includegraphics[width=.47\textwidth]{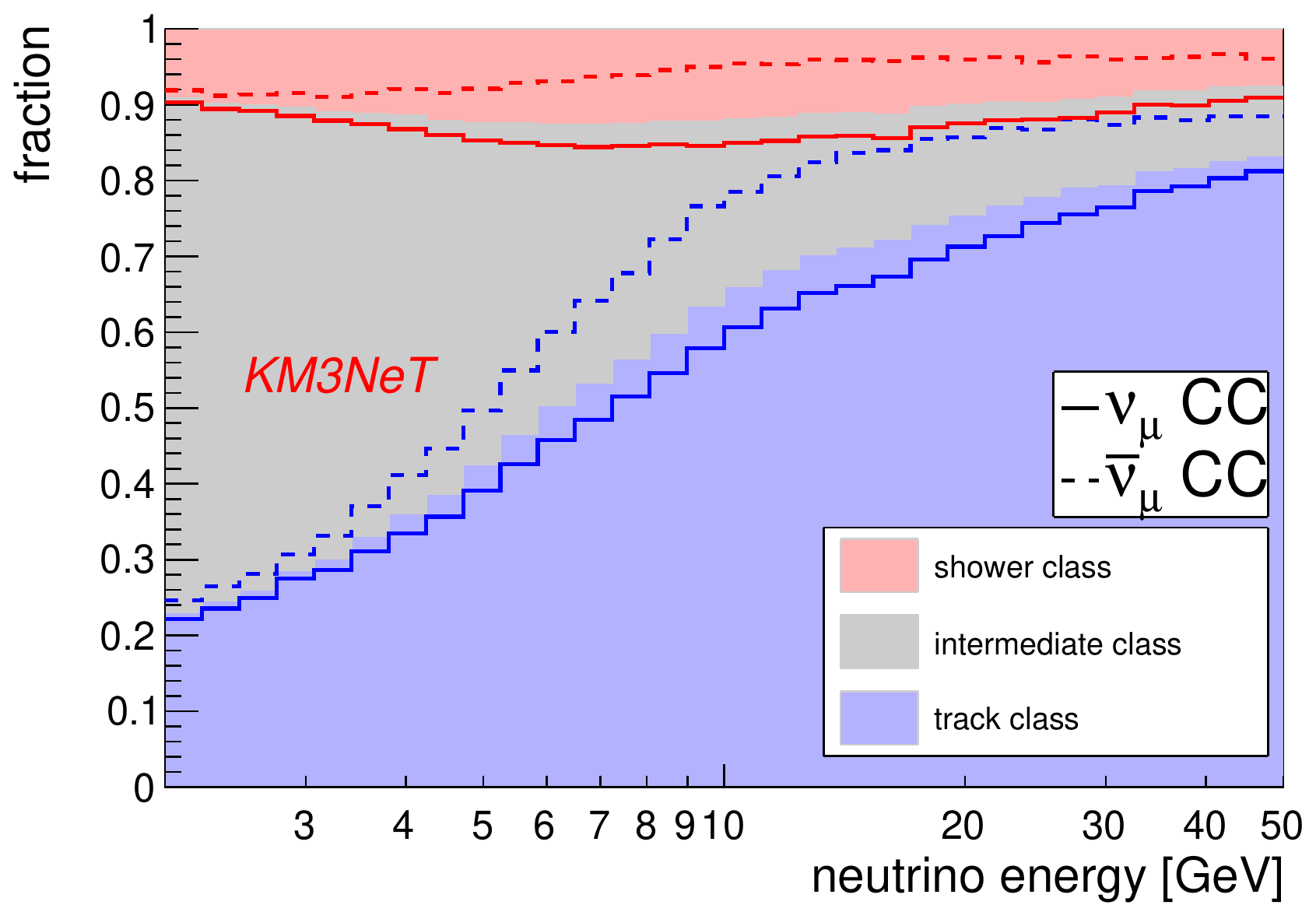}
    \hfill
    \includegraphics[width=.47\textwidth]{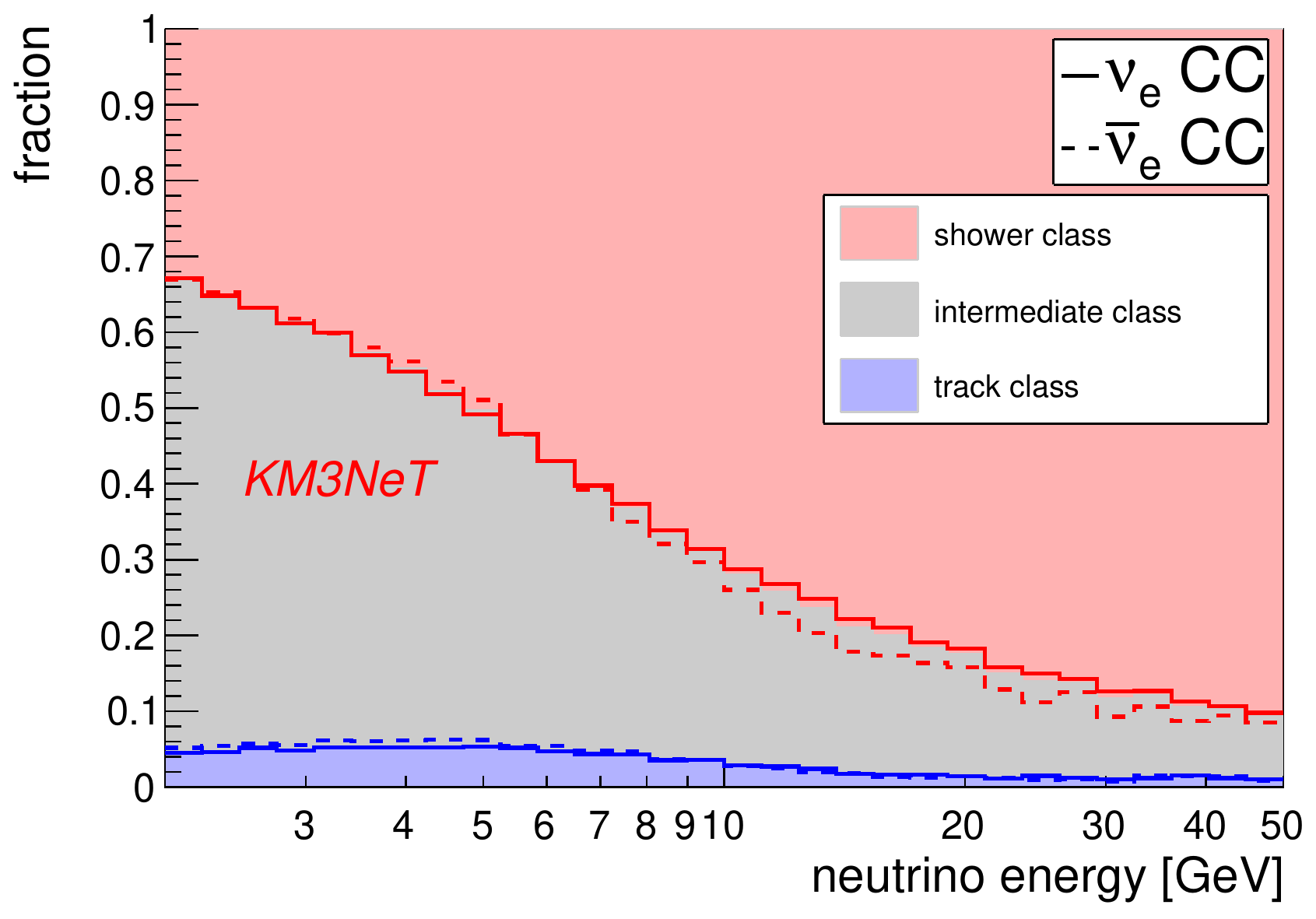}
    \includegraphics[width=.47\textwidth]{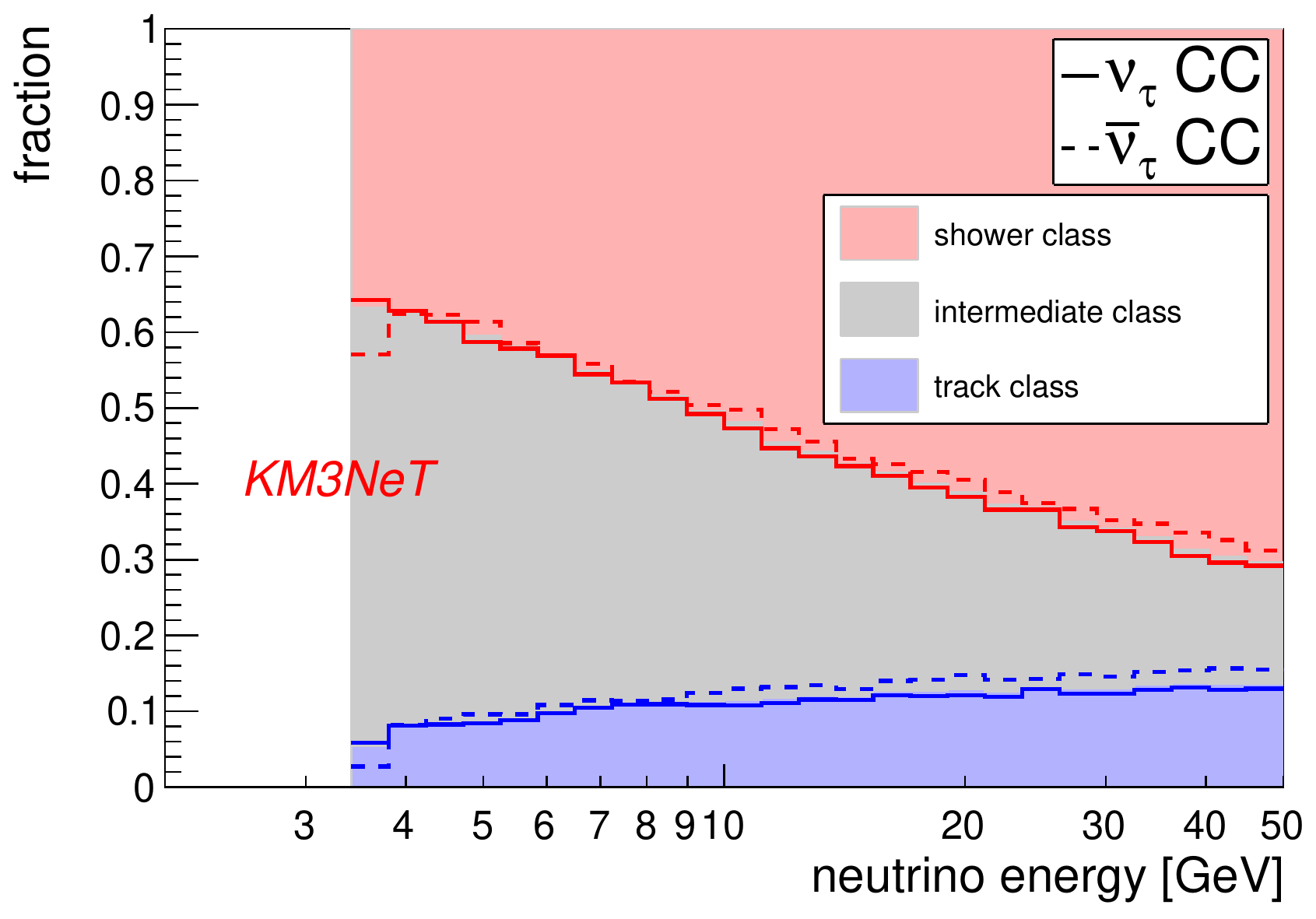}
    \hfill
    \includegraphics[width=.47\textwidth]{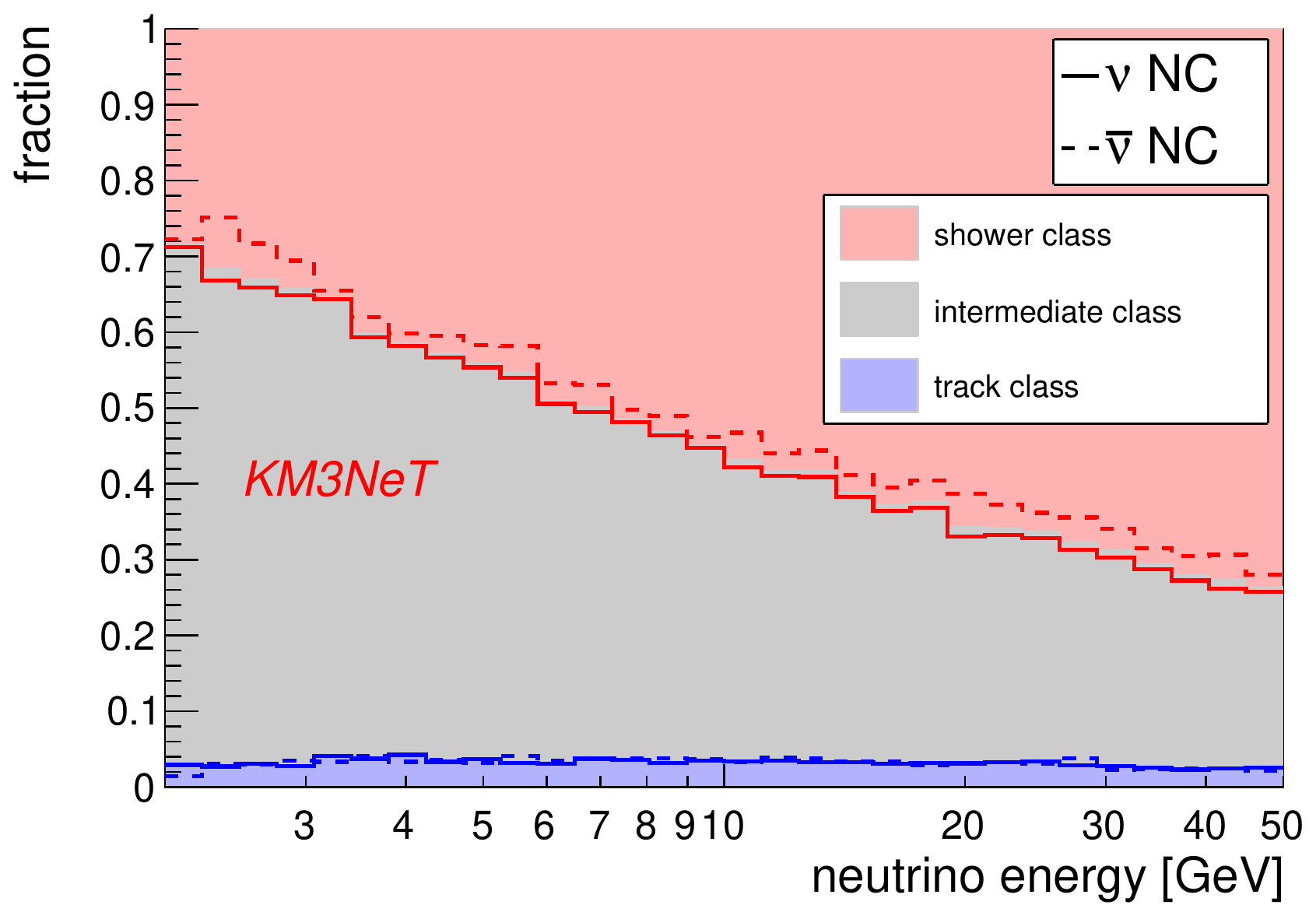}
    \caption{
    Fractions of preselected neutrino events of different types that are classified in the track class, the intermediate class, and the shower class, as a
    function of true neutrino energy. The definition of the classes is given in \autoref{eq:classification}. Coloured areas correspond to the composition of the atmospheric neutrino flux. Solid and dashed lines show individual fractions for neutrinos and anti-neutrinos, respectively.}
    \label{fig:track_selection}
\end{figure*}

\begin{figure}[htb]
    \centering
    \includegraphics[width=\columnwidth]{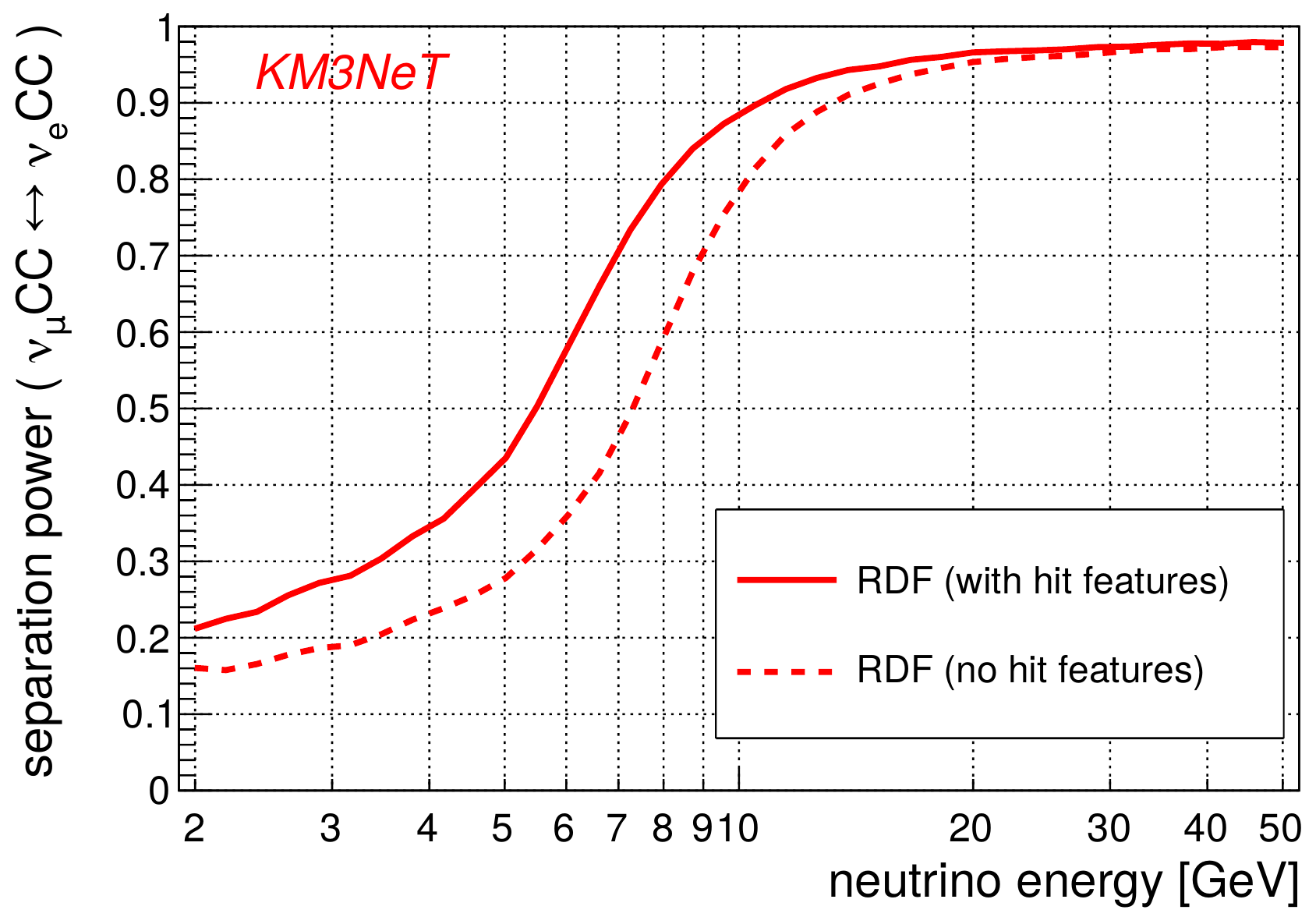}
    \caption{
    Comparison of the classifier performance as a function of true neutrino energy in terms of the separation power metric as defined in \autoref{eq:separability}. Separation power for training with (solid) and without (dashed) hit-based features is shown.}
    \label{fig:separability_plot}
\end{figure}

The \pc{performance} of the event type classifier for neutrinos is shown in \autoref{fig:track_selection}, where the fractions of events ending up in the respective class are presented as a function of neutrino energy.

\pc{The fraction of correctly classified events} increases steeply in the energy region up to $\sim\SI{15}{GeV}$, where less than $5\%$ of $\nuaneCC$ and $\nuanNC$ are mis-classified as tracks. At $\sim\SI{15}{GeV}$, 
85\% $\anumuCC$ and 70\% 
of $\numuCC$ are correctly classified as tracks. The better classification performance for $\anumuCC$ compared to $\numuCC$ is due to the different Bjorken-y distribution resulting in longer tracks of the final state muon for $\anumuCC$. The fraction of $\nuantauCC$ events classified as tracks is higher compared to $\nuaneCC$ and $\nuanNC$ reflecting the 17\% branching ratio for muonic tau decays.

To quantify the gain in classification performance when including the additional variables based on the expected hit distributions for $\nuanmuCC$ and $\nuaneCC$,
the separation power, ${S}$, is used. It quantifies the overlap in the distribution of the \texttt{track\_score} between $\nuanmuCC$ and $\nuaneCC$ events by using the correlation coefficient, ${C}$, and is defined as:
\begin{align}
\label{eq:separability}
{S}&(\Delta E) =  1 - {C}(\Delta E) =\notag \\ &1 - \frac{\sum_i P_{i,\text{score}}^{\nuanmu}(\Delta E) \cdot P_{i,\text{score}}^{\nuane}(\Delta E)}{\sqrt{\sum_i \left(P_{i,\text{score}}^{\nuanmu}(\Delta E)\right)^2\cdot\sum_i \left(P_{i,\text{score}}^{\nuane}(\Delta E)\right)^2}}.
\end{align}
The separation power is calculated in slices of neutrino energy $\Delta E$ by summing over binned probabilities for the \texttt{track\_score} values, $P_{i,\text{score}}$.
The resulting quantity is shown as a function of neutrino energy in \autoref{fig:separability_plot}.
The event type classification reaches 50\% separation power at 20\% lower neutrino energies when including hit-based variables in the classifier.

\section{Sensitivity Calculation}
\label{Sec:analysis}
\subsection{Method}\label{Sec:Method}

The neutrino oscillation parameters are studied by analysing the expected bi-dimensional distributions -- reconstructed energy, reconstructed cosine zenith angle -- of the neutrino candidates in the three event classes (track, intermediate and shower).

These distributions are obtained based on the true energy and cosine zenith angle event distributions split by neutrino interaction type ($\nueCC$, $\anueCC$ , $\numuCC$, $\anumuCC$,$\nutauCC$, $\anutauCC$, $\nuNC$, $\anuNC$). \pc{The true} distributions are derived from the neutrino flux \cite{Honda2015}, the neutrino cross section \cite{Zeller2012}, the probability for each neutrino flavour to oscillate while
traversing the Earth computed with the OscProb software \cite{OscProb} and a bi-dimensional parametric description of the detector effective volume. The latter is obtained based on the simulations described in  \autoref{Sec:Simulation}. 

Each of the eight true energy and cosine zenith angle distributions are then split in the three event classes (track, intermediate and shower), resulting in 24 distributions. The fractions of the distribution classified in each category, given the true neutrino energy, is obtained using parametric functions, derived from simulations.

The distributions of the reconstructed quantities are obtained from these 24 distributions using two sets of parametric functions that describe, first, the probability for a neutrino to be reconstructed at any energy given the true neutrino energy and, second, the probability for a neutrino to be reconstructed at any zenith angle given the true neutrino energy and true zenith angle.

These 24 distributions are merged to form the three final distributions of observables (reconstructed energy and cosine zenith angle) for events classified as track, intermediate and shower. 

These three final distributions are used as an Asimov data set~\cite{CowanEtAl_2011} to derive the median sensitivity to the oscillation parameters under study. A distribution obtained with a given set of oscillation parameters, the \textit{null} hypothesis, is confronted with other sets, the \textit{alternate} hypotheses, using $\lik_0$, the Poisson likelihood $\chi^2$ ~\cite{hep-ph_BakerEtAl_1984}, defined as:

\begin{align}
	\lik_0 & = \sum_{ i \in \rm{[E^{rec},~cos\theta_z^{rec}]}} \lik_{0,i}\nonumber                                                                            \\
	         & = \sum_{ i \in \rm{[E^{rec},~cos\theta_z^{rec}]}}  -2.0 \cdot (n_i^{\rm{null}}-n_i^{\rm{alt}} -  n_i^{\rm{null}}\ln \frac{n_i^{\rm{null}}}{n_i^{\rm{alt}}}), 
	\label{eq:chi2}
\end{align}

where $n^{\rm{null}}_i$ and $n^{\rm{alt}}_i$ are the expected numbers of events under the \textit{null} and \textit{alternate} hypotheses, respectively, in the $i^{th}$ region of the reconstructed energy -- cosine zenith angle plane.

Relevant external information on the neutrino oscillation parameters~\cite{globalFitEsteban} and model uncertainties are taken into account by adding to $\lik_0$ extra contributions measuring the discrepancy between the parameter value, $p_i^{obs}$, and the one expected, $p_i^{exp}$, in standard deviation unit, $\sigma_i$:
\begin{align}
  \lik_{\rm{eff}} = \lik_0 + \sum_{ i \in \rm{parameters}}  \frac{ (p_i^{\rm{exp}} - p_i^{\rm{obs}})^2}{\sigma_i^2}. 
\end{align}
The sensitivity to the parameters under study (described in the next sections) is obtained from the $\lik_{\rm{eff}}$, minimised over all remaining parameters, as $\sqrt{\lik_{\rm{eff},\rm{min}}}$.

A first set of model parameters reflecting the current knowledge on the neutrino flux are considered using the uncertainties reported in ~\cite{BarrEtAl_2006}:
\begin{enumerate}
	\item the \textit{spectral index} of the neutrino flux energy distribution is allowed to vary without constraint,
	\item \label{sysnew-1} the ratio of upgoing to horizontally-going neutrinos, $n_{\nuan_{up}} / n_{\nuan_{horiz}}$ , is allowed to vary with a standard deviation of 2\% of the parameter's nominal value,
	      
	\item the ratio between the total number of $\nuane$ and $\nuanmu$, $n_{\nuane} / n_{\nuanmu}$, is allowed to vary  with a standard deviation of 2\% of the parameter's nominal value,
	        
	\item \label{sysnew-2} the ratio between the total number of $\nue$ and $\anue$, $n_{\nue} / n_{\anue}$, is allowed to vary with a standard deviation of 7\% of the parameter's nominal value,
	        
	\item the ratio between the total number of  $\numu$ and $\anumu$, $n_{\numu} / n_{\anumu}$, is allowed to vary with a standard deviation of 5\% of the parameter's nominal value.
\end{enumerate}

In addition, two uncertainties on the neutrino cross section are considered:
\begin{enumerate}
	\setcounter{enumi}{5}
	\item the number of NC events is scaled by a factor $n_{NC}$ to which no constraint is applied,
	\item the number of $\nuantauCC$ is scaled by a factor $n_\tau^{CC}$ to which no constraint is applied.
\end{enumerate}

\begin{table*}[p]

  \sisetup{
		table-align-uncertainty=true,
		separate-uncertainty=true,
	}
				
	\caption{Parameter values minimising the \lik\ obtained for three years of data taking with NO (IO) as \textit{null} hypothesis and IO (NO) as \textit{alternate} hypothesis and using the oscillation parameters from~\autoref{tab:ParamNMH}. The parameter uncertainties are defined as the values by which the parameter has to vary to increase \lik\ by 1.0. For each parameter value scanned, \lik\ is minimised over the other free parameters. 
	}
	\label{tab:FitResults}
	\begin{center}
		\newcommand{\lab}[1]{%
			\multirow{2}{*}{#1}
		}
		\renewcommand{\arraystretch}{1.2}
		\footnotesize
		\begin{tabular}{cccccc}\hline
																	
			Parameter                                  & Null hypothesis & Dataset Value       & Value at Min.                              & Prior       \\\hline\hline
			\lab{$\dmsq{32}$ [\si{eV^2}]}              & NO              & \num{2.528d-3}      & $(2.51 ^{+ 0.11 }_{-0.11})\times 10^{-3}$  & \lab{free}  \\
			                                           & IO              & \num{2.436d-3}      & $(2.43 ^{+ 0.10 }_{-0.08})\times 10^{-3} $ &             \\\hline
			\lab{$\dmsq{21}$ [\si{eV^2}]}              & NO              & \lab{\num{7.39d-5}} & \lab{\num{7.39d-5}}                        & \lab{fixed} \\
			                                           & IO              &                     &                                            &             \\\hline
			\lab{$\deltaCP$  [\si{\degree}]}           & NO              & 221.0               & $162 \pm 180$                              & \lab{free}  \\
			                                           & IO              & 282.0               & $190 \pm 180$                              &             \\\hline
			\lab{$\theta_{13}$ [\si{\degree}]}         & NO              & 8.60                & $8.63 \pm{0.40}$                           & \lab{0.13}  \\
			                                           & IO              & 8.64                & $8.62 \pm{0.29}$                           &             \\\hline
			\lab{$\theta_{12}$ [\si{\degree}]}         & NO              & \lab{33.82}         & \lab{33.82}                                & \lab{fixed} \\
			                                           & IO              &                     &                                            &             \\\hline
			\lab{$\theta_{23}$ [\si{\degree}]}         & NO              & 48.6                & $  49.4 ^{+ 2.3 }_{-3.9}$                  & \lab{free}  \\
			                                           & IO              & 48.8                & $  41.5 ^{+ 3.8 }_{-1.9}$                  &             \\\hline
			\lab{Spectral index}                       & NO              & \lab{1.0}           & $  1.00 \pm 0.02 $                         & \lab{free}  \\
			                                           & IO              &                     & $  1.01 \pm 0.02 $                         &             \\\hline
			\lab{$n_{\nuan_{up}} / n_{\nuan_{horiz}}$} & NO              & \lab{1.0}           & $  1.01 \pm 0.01 $                         & \lab{0.02}  \\
			                                           & IO              &                     & $  1.00 \pm 0.01 $                         &             \\\hline								   
			\lab{$n_{\nuane} / n_{\nuanmu}$}           & NO              & \lab{1.0}           & $  1.02 \pm 0.06 $                         & \lab{0.02}  \\
			                                           & IO              &                     & $  1.00 ^{+0.05 }_{-0.04}$                 &             \\\hline
			\lab{$n_{\nue} / n_{\anue}$}               & NO              & \lab{1.0}           & $  1.02 ^{+0.22 }_{-0.21}$                 & \lab{0.07}  \\
			                                           & IO              &                     & $  1.00 \pm 0.16 $                 	&             \\\hline
			\lab{$n_{\numu} / n_{\anumu}$}             & NO              & \lab{1.0}           & $  0.98 ^{+0.15 }_{-0.14}$                 & \lab{0.05}  \\
			                                           & IO              &                     & $  1.00 ^{+0.12 }_{-0.11}$                 &             \\\hline
			\lab{Energy scale}                         & NO              & \lab{1.0}           & $  1.02 \pm 0.05 $                         & \lab{0.06}  \\
			                                           & IO              &                     & $  0.99 \pm 0.04 $                         &             \\\hline
			\lab{Had. energy scale}                    & NO              & \lab{1.0}           & $  0.96 ^{+0.13 }_{-0.10}$                 & \lab{0.05}  \\
			                                           & IO              &                     & $  1.00 ^{+0.11 }_{-0.08}$                 &             \\\hline
			\lab{$n_{NC}$}                             & NO              & \lab{1.0}           & $  1.02 ^{+0.42 }_{-0.37}$                 & \lab{free}  \\
			                                           & IO              &                     & $  0.89 ^{+0.32 }_{-0.28}$                 &             \\\hline
			\lab{$n_\tau^{CC}$}                        & NO              & \lab{1.0}           & $  1.05 ^{+0.19 }_{-0.20}$                 & \lab{free}  \\
			                                           & IO              &                     & $  1.03 ^{+0.13 }_{-0.14}$                 &             \\\hline
			\lab{$n_{Intermediate}$ }                        & NO              & \lab{1.0}           & $  1.00 ^{+0.05 }_{-0.06}$                 & \lab{free}  \\
			                                           & IO              &                     & $  1.02 \pm 0.04 $                         &             \\\hline
			\lab{$n_{Tracks}$}                         & NO              & \lab{1.0}           & $  0.98 \pm 0.04 $                         & \lab{free}  \\
			                                           & IO              &                     & $  1.00 \pm 0.03 $                         &             \\\hline
			\lab{$n_{Showers}$}                        & NO              & \lab{1.0}           & $  1.01 ^{+0.09 }_{-0.08}$                 & \lab{free}  \\
			                                           & IO              &                     & $  1.03 ^{+0.07 }_{-0.06}$                 &             \\\hline\hline
																					
		\end{tabular}
	\end{center}
\end{table*}

Then three uncertainties on the detector response are taken into account:
\begin{enumerate}
	\setcounter{enumi}{7}
	\item the absolute \textit{energy scale} of the detector depends on the knowledge of the PMT efficiencies and the water optical properties, as shown in \cite{LoI} (section 3.4.6). The time dependent PMT efficiencies are monitored permanently with high fidelity, using coincidence signals from $^{40}$K decays, as demonstrated in ANTARES \cite{ANTARES_AlbertEtAl_2018}. Several methods are under study to monitor in-situ the water optical properties, exploiting both Cherenkov light from atmospheric muons and $^{40}$K decays as well as signals from artificial light sources. The combination of these methods will allow to constrain the energy scale uncertainty to a few percent. In the study presented here, the energy scale of the detector is allowed to vary with a standard deviation of 5\% around its nominal value,
	         
	\item the light yield in hadronic showers, \textit{Had. Energy Scale} is allowed to vary  with a standard deviation of 6\% of the parameter's nominal value, as obtained while comparing two different simulation software packages Gheisha and Fluka~\cite{IntrinsicPaper},
	\item  \label{sysnew-L} the number of events in the three classes is allowed to vary without constraints via three scaling factors $n_{\rm{Tracks}}$, $n_{\rm{Intermediate}}$, $n_{\rm{Showers}}$.
	      
\end{enumerate}
Previous studies \cite{LoI,premModel} showed that the uncertainty on the Earth model had negligible effects on the NMO sensitivity and is thus ignored in this study. 
Systematics \ref{sysnew-1} and \ref{sysnew-2}--\ref{sysnew-L} were not included in the previous analysis~\cite{LoI}.
\autoref{tab:FitResults} reports all the parameters and the external constraints applied to them.

\subsection{NMO Sensitivity}

The sensitivity to the neutrino mass ordering is obtained as a function of $\theta_{23}$ using the method described in \autoref{Sec:Method}. For every $\theta_{23}$ value, each mass ordering hypothesis -- the \textit{null} hypothesis -- is confronted with the reversed one -- the \textit{alternate} hypothesis. The oscillation parameters used for the \textit{null} hypothesis are reported in \autoref{tab:ParamNMH} as well as the constraints applied to them in the minimisation procedure.

\begin{table}[!ht]
  \caption{Oscillation parameters values used for different analyses for the \textit{null} hypothesis and constraints applied during the $\lik_{eff}$ minimisation. The values are taken from~\cite{globalFitEsteban} except the ones identified by a dagger ($\dagger$) which are extra $\theta_{23}$ and $\deltaCP$ test points used for the NMO sensitivity.}
	\label{tab:ParamNMH}
	\begin{center}
		\newcommand{\idems}[1]{%
			\multicolumn{2}{c|}{#1}
		}
		\newcommand{\lab}[1]{%
			\multirow{2}{*}{#1}
		}
	\resizebox{\columnwidth}{!}{
		\begin{tabular}{cccc}\hline
			Parameter           &    & Null Hypothesis Values            & Constraints         \\\hline\hline 
			$\dmsq{21}$         &    & \num{7.39d-5} \si{eV^2}           & fixed               \\\hline 
			$\theta_{12}$       &    & \ang{33.82}                       & fixed               \\\hline 
			\lab{$\theta_{13}$} & NO & \ang{8.60}                        & \lab{\ang{\pm0.13}} \\
			                    & IO & \ang{8.64}                        &                     \\\hline 
			\lab{$\dmsq{31}$}   & NO & \num{2.528d-3} \si{eV^2}          & \lab{free}          \\
			                    & IO & \num{2.436d-3} \si{eV^2}          &                     \\\hline 
			\lab{$\theta_{23}$} & NO & \ang{48.6}, [\ang{40}--\ang{50}]$\dagger$  & \lab{free}          \\
			                    & IO & \ang{48.8}, [\ang{40}--\ang{50}]$\dagger$  &                     \\  \hline 
			\lab{$\deltaCP$}    & NO & \ang{221.0}, \ang{0}$\dagger$, \ang{180.0}$\dagger$ & \lab{free}          \\
			                    & IO & \ang{282.0}, \ang{0}$\dagger$, \ang{180.0}$\dagger$ &                     \\\hline 
		\end{tabular}
		}
	\end{center}
\end{table}

\begin{figure*}[!htb]
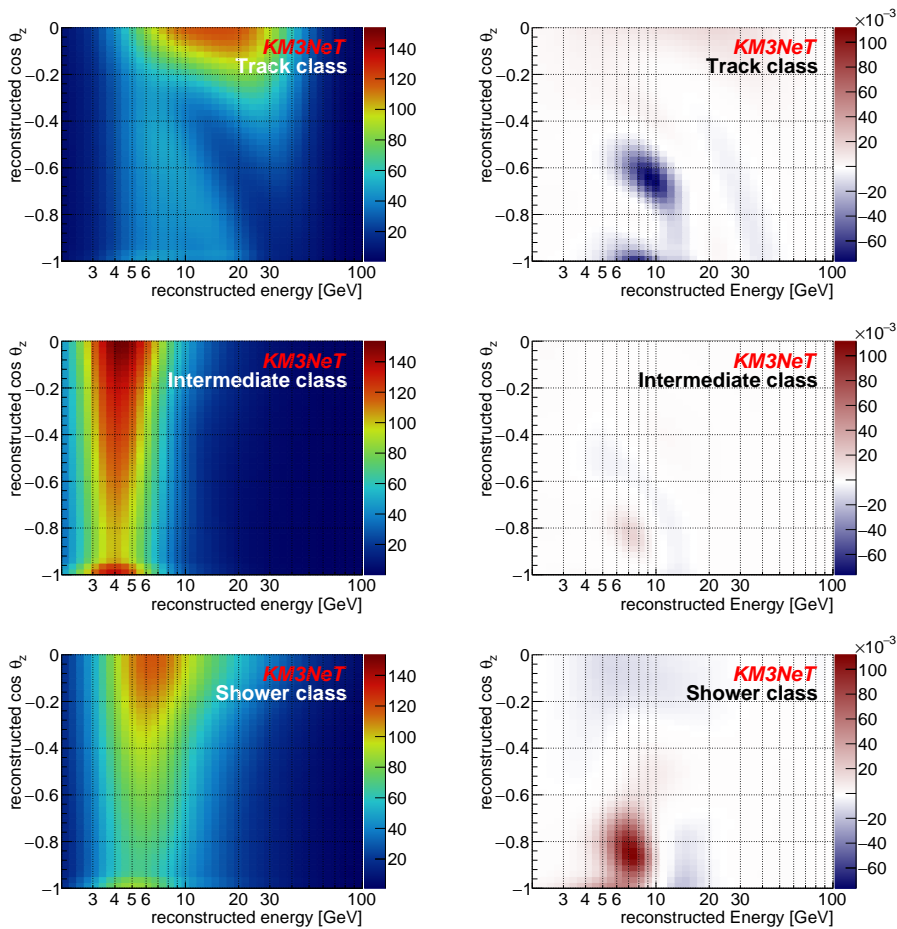

	\centering
	
	\includegraphics[width=0.8\columnwidth]{{../plots/sensitivity/NO-NuFit41-3y_Tracks-NULL}.pdf}
	\includegraphics[width=0.8\columnwidth]{{../plots/sensitivity/NO-NuFit41-3y_Tracks-Chi2}.pdf}\\
		
	\includegraphics[width=0.8\columnwidth]{{../plots/sensitivity/NO-NuFit41-3y_Middle-NULL}.pdf}
	\includegraphics[width=0.8\columnwidth]{{../plots/sensitivity/NO-NuFit41-3y_Middle-Chi2}.pdf}\\
		
	\includegraphics[width=0.8\columnwidth]{{../plots/sensitivity/NO-NuFit41-3y_Showers-NULL}.pdf}
	\includegraphics[width=0.8\columnwidth]{{../plots/sensitivity/NO-NuFit41-3y_Showers-Chi2}.pdf}\\
	
	\caption{ (left) Expected event distributions for NO after 3 years of data taking for events classified as track (top), intermediate (middle), and shower (bottom). \pc{(right) Signed binned Poisson likelihood $\chi^2$ derived using these distributions and the ones obtained minimising $\lik_{\rm{eff}}$ with the IO hypothesis. If more events are expected for NO than for IO, the value plotted is $\lik_{0,i}$ which, as defined in \autoref{eq:chi2}, is positive. Otherwise, the value plotted is $-\lik_{0,i}$.}}
	\label{fig:EvtDistribution}
\end{figure*}

The distributions of selected events after three years of data taking for the \textit{null} hypothesis assuming NO, $n^{\rm{null}}_i$, obtained with the parametric detector response are shown in \autoref{fig:EvtDistribution} using a 40$\times$40 grid of energy, equally logarithmically spaced between 2 and \SI{100}{GeV}, and cosine zenith angle equally spaced between $0$ and $-1$. Around \num{51d3} events are expected for the track-class, \num{63d3} for the intermediate-class and \num{64d3} for the shower-class. \autoref{fig:EvtDistribution} shows also the $\lik_{0, i,\rm{min}}$ obtained confronting these distributions with the \textit{alternate} hypothesis ones.

\begin{figure}[ht]
  \centering
	\subfigure[]{
		
\includegraphics[width=\columnwidth]{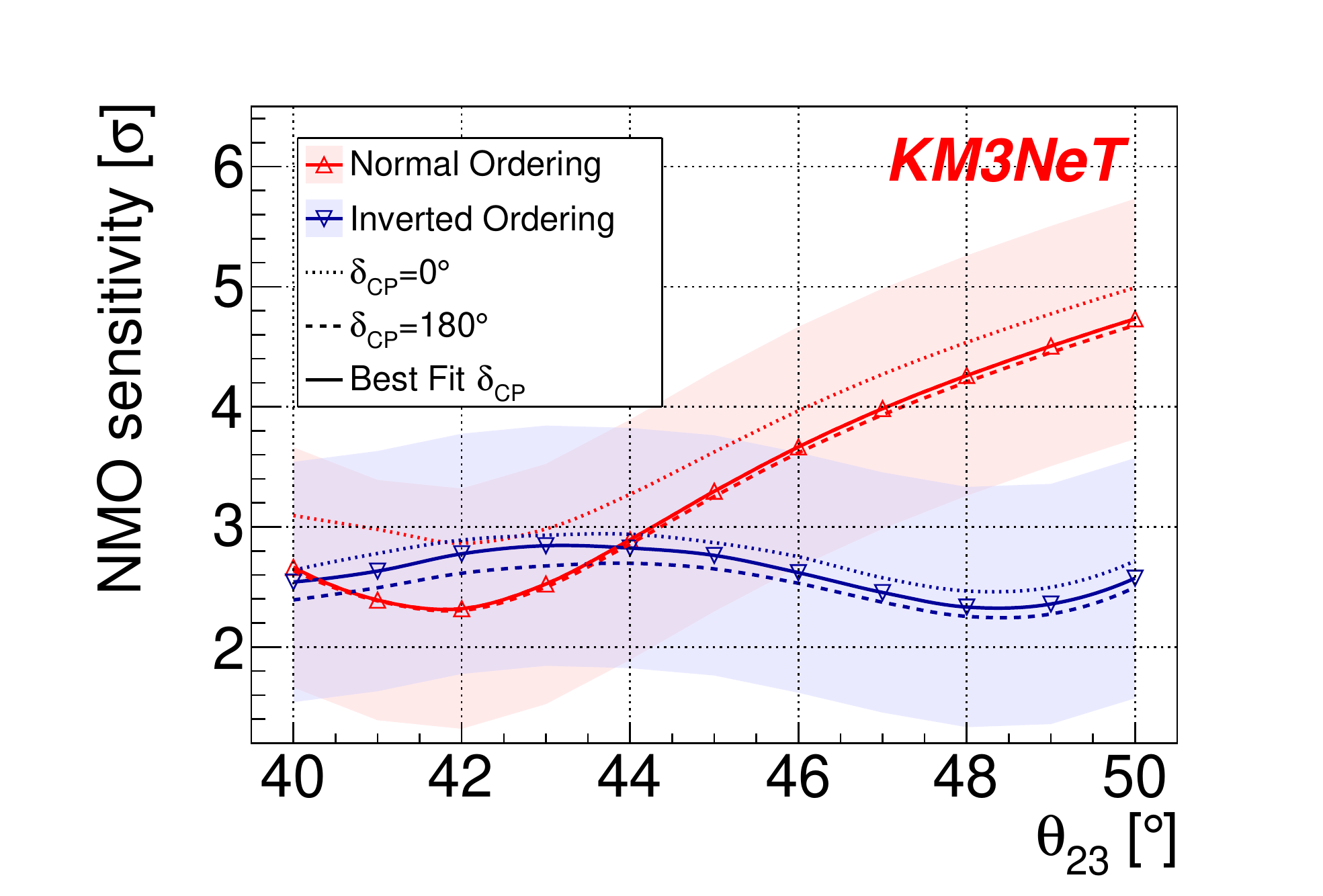}
		\label{fig:SensitivityNMHTh23}
	}\vspace{-3mm}
	\subfigure[]{	
\includegraphics[width=\columnwidth]{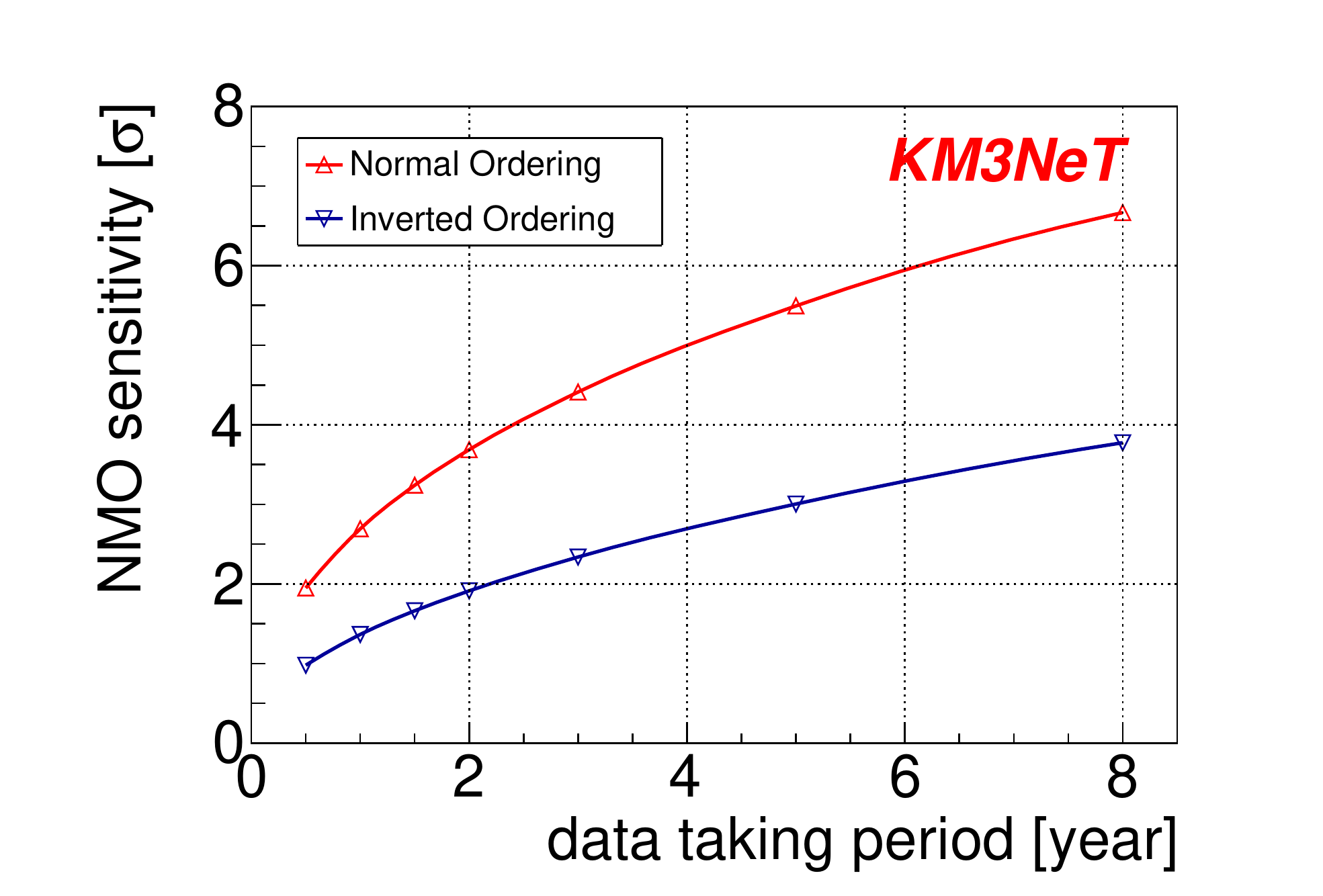}
		\label{fig:SensitivityNMHTime}
	}
	\caption{ \protect\subref{fig:SensitivityNMHTh23} Sensitivity to NMO after three years of data taking, as a function of the true $\theta_{23}$ value, for both normal (red upward pointing triangles) and inverted ordering (blue downward pointing triangles) under three assumptions for the $\deltaCP$ value: the world best fit point for NO, IO reported in \autoref{tab:ParamNMH} (plain line), \ang{0} (dotted line) or \ang{180} (dashed line).  The coloured shaded areas represent the sensitivity that 68\% of the experiment realisation would yield, according to the Asimov approach~\cite{CowanEtAl_2011}. \protect\subref{fig:SensitivityNMHTime}  Sensitivity to NMO as a function of data taking time for both normal (red upward pointing triangles) and inverted ordering (blue downward pointing triangles) and assuming the oscillation parameters reported in \autoref{tab:ParamNMH}. }
	\label{cc}
\end{figure}

The sensitivity to the NMO after three years of data taking is reported as a function of $\theta_{23}$ for both NMO in \autoref{fig:SensitivityNMHTh23}. Assuming the  current best estimates for $\theta_{23}$ (see \autoref{tab:ParamNMH}), the NMO sensitivity is \nmoNO$\sigma$ if the true NMO is NO and \nmoIO$\sigma$ if it is IO. 
\autoref{tab:FitResults} illustrates the fit results at one test point for oscillation parameters reported in \autoref{tab:ParamNMH}. None of the systematic uncertainties exhibits a strong pull in this wrong-hierarchy fit, demonstrating that degeneracies between the NMO choice and systematic uncertainties are generally small.

\autoref{fig:SensitivityNMHTime} shows the sensitivity for both NMO as a function of data taking time. The NMO can be determined at 3$\sigma$ level after \yNO\ years if the true NMO is NO, and after \yIO\ years if it is IO.
 
\subsection{Sensitivity to $\dmsq{32}$ and $\theta_{23}$}

\begin{figure}[!ht]
  \centering
	\subfigure[]{
		
\includegraphics[width=\columnwidth]{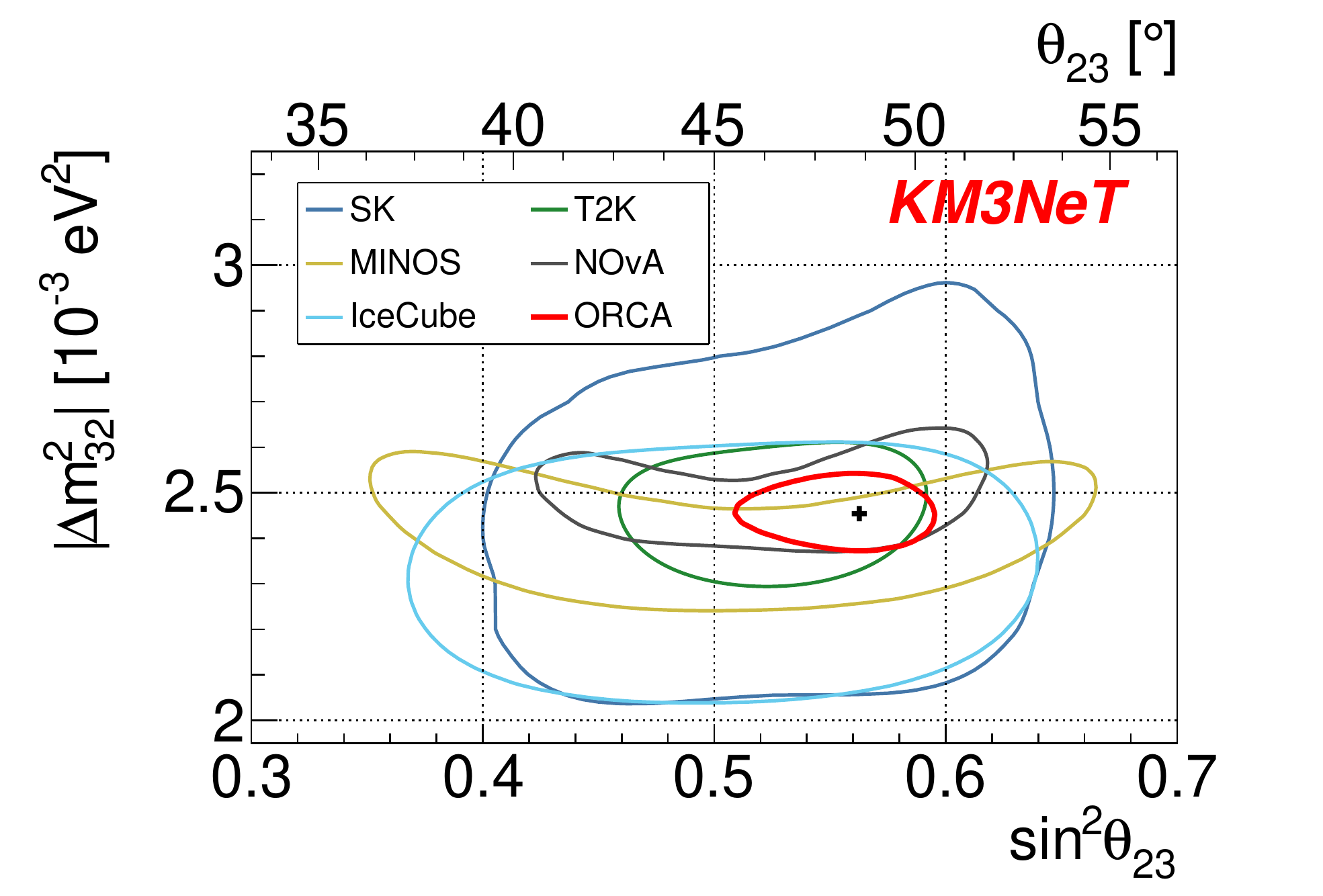}
		\label{fig:ContourNOAll}
	}\vspace{-3mm}
	\subfigure[]{
		
\includegraphics[width=\columnwidth]{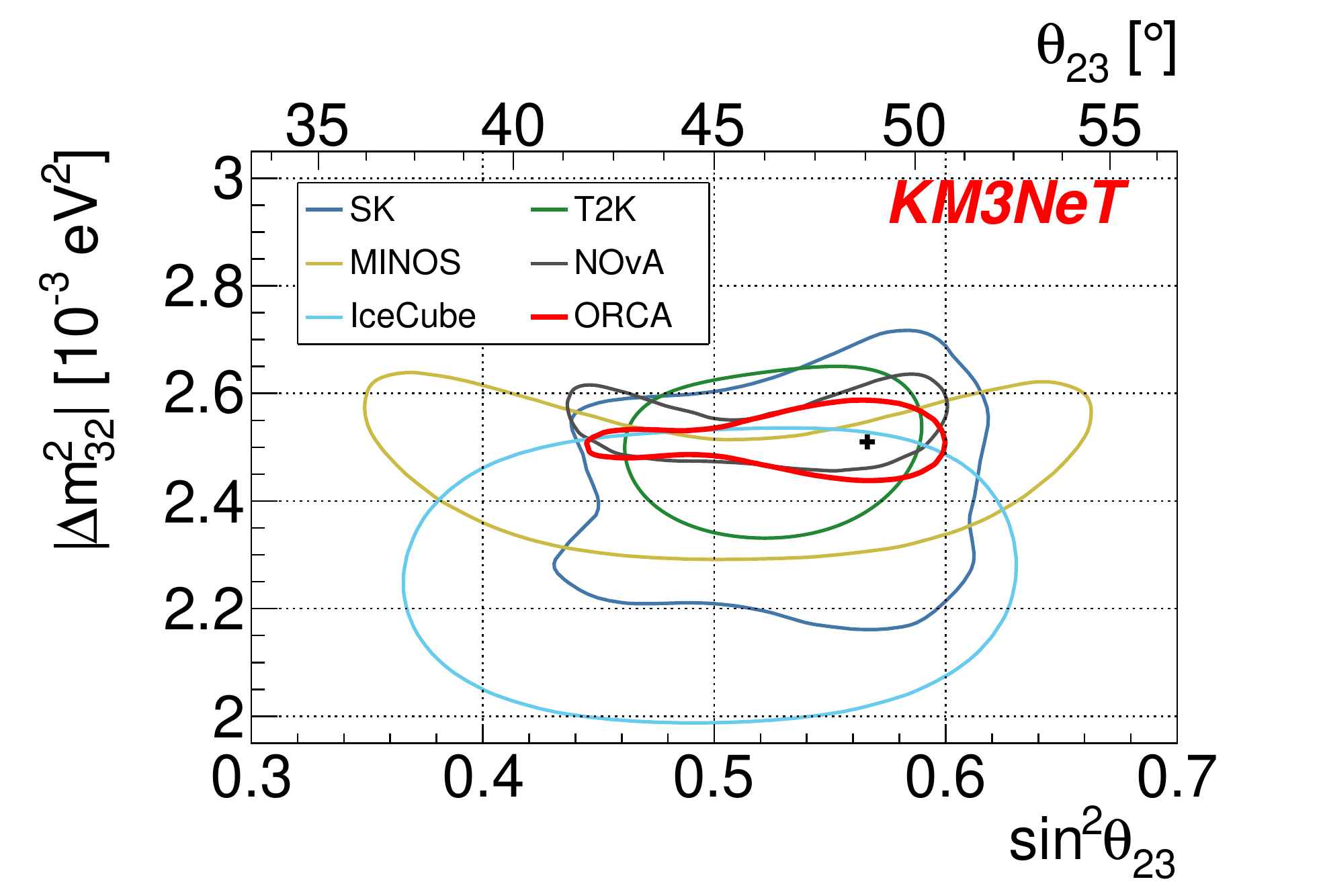}
		\label{fig:ContourIOAll}
	}
	\caption{  Expected measurement precision of $\dmsq{32}$ and $\theta_{23}$ for both NO \protect\subref{fig:ContourNOAll} and IO \protect\subref{fig:ContourIOAll} after 3 years of data taking at 90\% confidence level (red) overlaid with results from other experiments~\cite{DeepCore2017,hep-ph_SuperKamiokande_2018,hep-ph_T2K_AbeEtAl_2020,aurisano_adam_2018_1286760,hep-ph_NOvA_AceroEtAl_2019} and the oscillation parameters reported in \autoref{tab:ParamNMH} (black cross).}
	\label{fig:contour}
\end{figure}

The sensitivity to $\dmsq{32}$ and $\theta_{23}$ is obtained using the method described in \autoref{Sec:Method}. The \textit{null} hypothesis, assuming the latest oscillation parameter values, reported in \autoref{tab:ParamNMH}, is confronted with a set of \textit{alternate} hypotheses, one for each point in the $\dmsq{32}$, $\theta_{23}$ plane. The NMO is kept fixed in the $\lik_{\rm{eff}}$ minimisation.
All ($\dmsq{32}$, $\theta_{23}$ ) points for which the resulting $\lik_{\rm{eff,min}}$ exceeds by 4.61~\cite{PDG2018} the $\lik_{\rm{eff}}$ minimum in the ($\dmsq{32}$, $\theta_{23}$ ) plane are excluded with 90\% confidence level. The oscillation parameters used and the constraints applied during the $\lik_{\rm{eff}}$ minimisation are reported in \autoref{tab:ParamNMH}. The resulting 90\% confidence level contours for both NMO are shown in \autoref{fig:contour}. The 90\% confidence level interval on $\dmsq{32}$ and $\theta_{23}$ are \dmErrNO\ and \dThErrNO\ for NO and, \dmErrIO\ and \dThErrIO\ for IO.

\begin{figure}[ht]
  \centering
	\subfigure[]{
\includegraphics[width=\columnwidth]{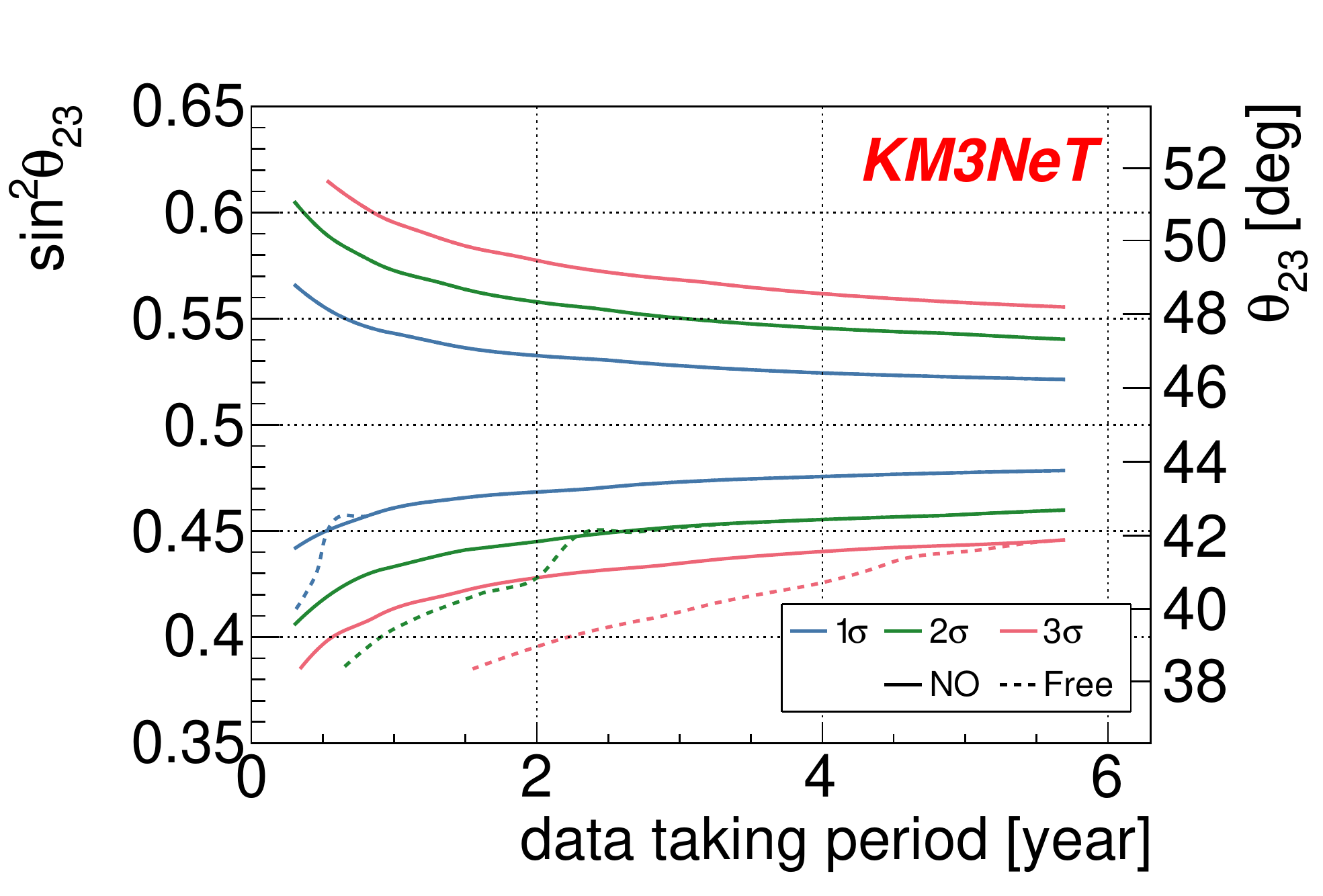}
		\label{fig:octantNO}
	}\vspace{-3mm}
	\subfigure[]{
\includegraphics[width=\columnwidth]{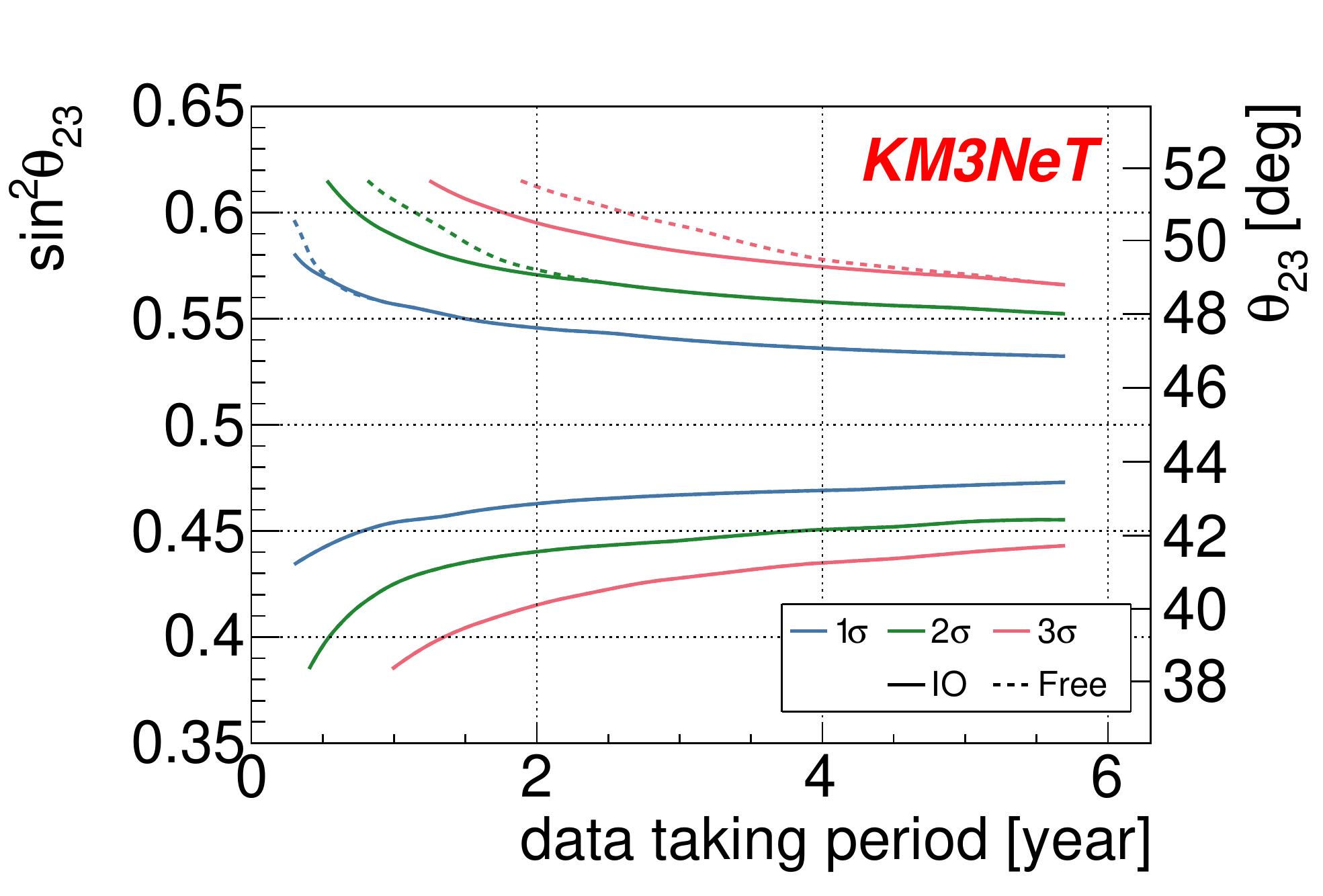}
		\label{fig:octantIO}
	}
	\caption{  Expected sensitivity to determine the $\theta_{23}$ octant at 1 (blue), 2 (green) or 3$\sigma$ (red) as a function of data taking time for both NO \protect\subref{fig:octantNO} and IO \protect\subref{fig:octantIO} assuming the true NMO is known (solid line) or unknown (dashed line). The dashed lines differ from the plain ones when the $\lik_{\rm eff}$ minimisation converges to the wrong NMO.}
	\label{fig:octant}
\end{figure}

The same analysis allows to calculate the significance to determine the 
octant of $\theta_{23}$. The \textit{alternate} hypothesis is now the minimal $\lik_{\rm eff}$ for 
$\theta_{23}$ in the opposite octant with respect to the true $\theta_{23}$ value. The 
results are shown in \autoref{fig:octant}, which illustrates the needed data taking 
time to reach a 1, 2 and 3$\sigma$ octant significance as a function of the 
true value of $\theta_{23}$. Dashed lines ignore the NMO, while for solid lines 
the NMO is assumed to be known.
\pc{KM3NeT/ORCA can constrain the octant with better than 95\% confidence level after 6 years of data taking for $\left|\sinsq{23}-0.5\right| < 0.05$.}

\subsection{Sensitivity to $\nuantau$ appearance}
The appearance of $\nuantau$ is determined by measuring the normalisation factor $n_{\nuantau}$ of the $\nuantau$ contribution. For this study, NO is assumed. As in the analyses above, the oscillation parameter values are taken from \autoref{tab:ParamNMH} and the normalisation is fixed to $n_{\nuantau} \equiv 1$ for the \textit{null} hypothesis. The latter is expected if the commonly accepted picture of unitary $3\times3$ neutrino mixing is complete and, in addition, the assumed standard model cross sections are correct. A measurement in tension with $n_{\nuantau} \equiv 1$ would therefore provide a model-independent test for new physics. Two choices to scale the $\nuantau$ contribution are possible for the \textit{alternate} hypotheses. The first is to vary only the $\nuantau$ CC contribution, leaving the NC contribution fixed to unity. The second allows for a combined CC+NC scaling of the $\nuantau$ flux.
Note, that the CC-only case correlates directly with a scaling of the $\nuantau$ CC cross section. Both choices, CC-only and CC+NC normalisation scaling, have been adopted in previous experiments (\cite{tau_appearance_superk_2017,tau_appearance_opera_finalresults2018} and \cite{tau_appearance_icecube_2019}, respectively).

The sensitivity is evaluated using the method described in \autoref{Sec:Method} extended by the additional scaling parameter $n_{\nuantau}$, affecting the $\nuantau$ CC flux and in case of CC + NC scaling also the NC fraction that has oscillated into the $\nuantau$ channel. 
While oscillations of the NC do not need to be considered if the overall flux remains unchanged, this is different for $n_{\nuantau} \neq 1$. In this case the procedure to populate the event distributions is modified and includes the oscillated fractions of each flavour, which allows to scale the $\nuantau$ contribution accordingly.

\begin{figure}[!ht]
	\centering
	\subfigure[]{ 
	  \includegraphics[width=.37\textwidth]{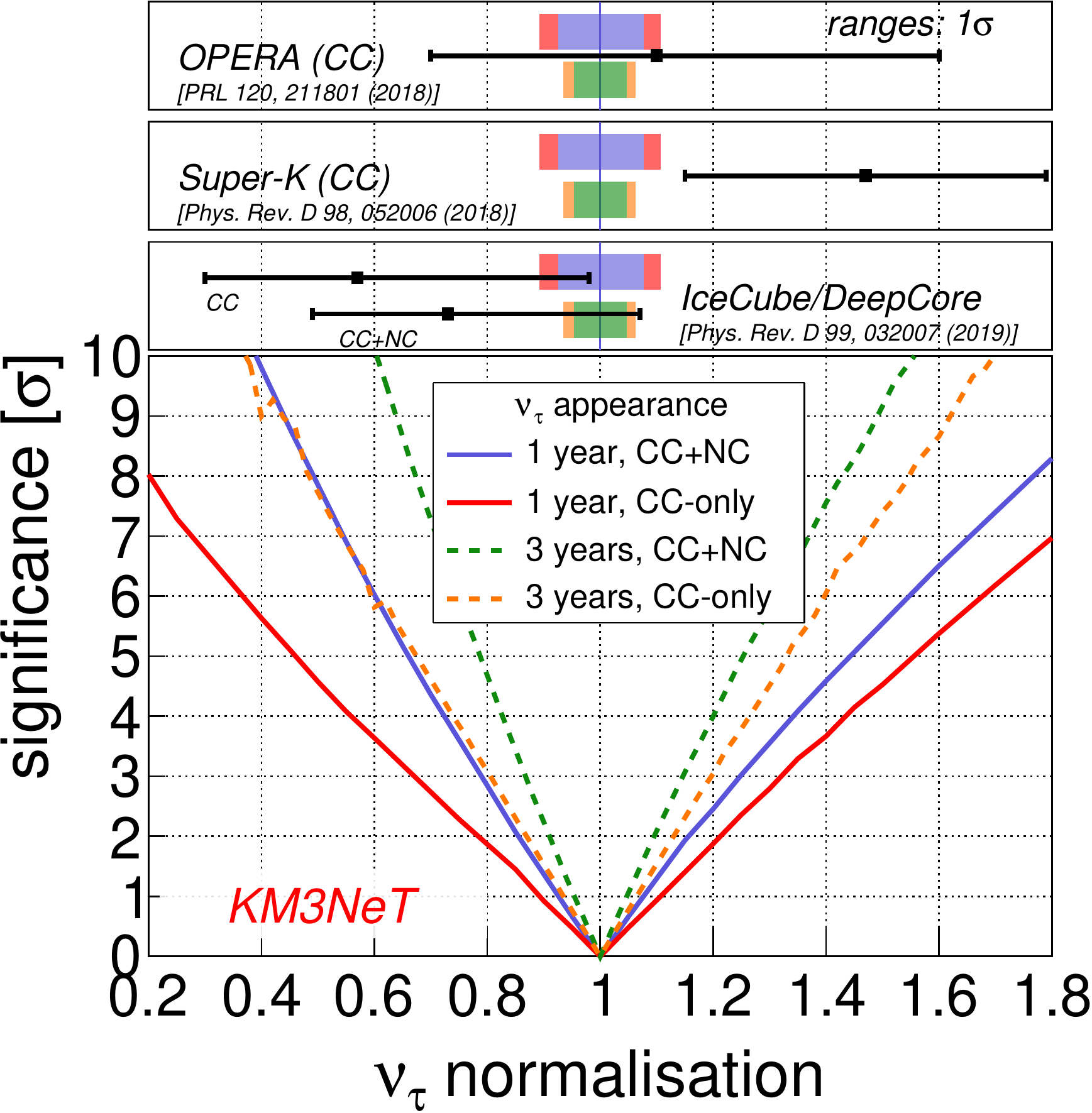}
		\label{fig:tauappearance1D}
	}
	\subfigure[]{ 
		\includegraphics[width=.45\textwidth]{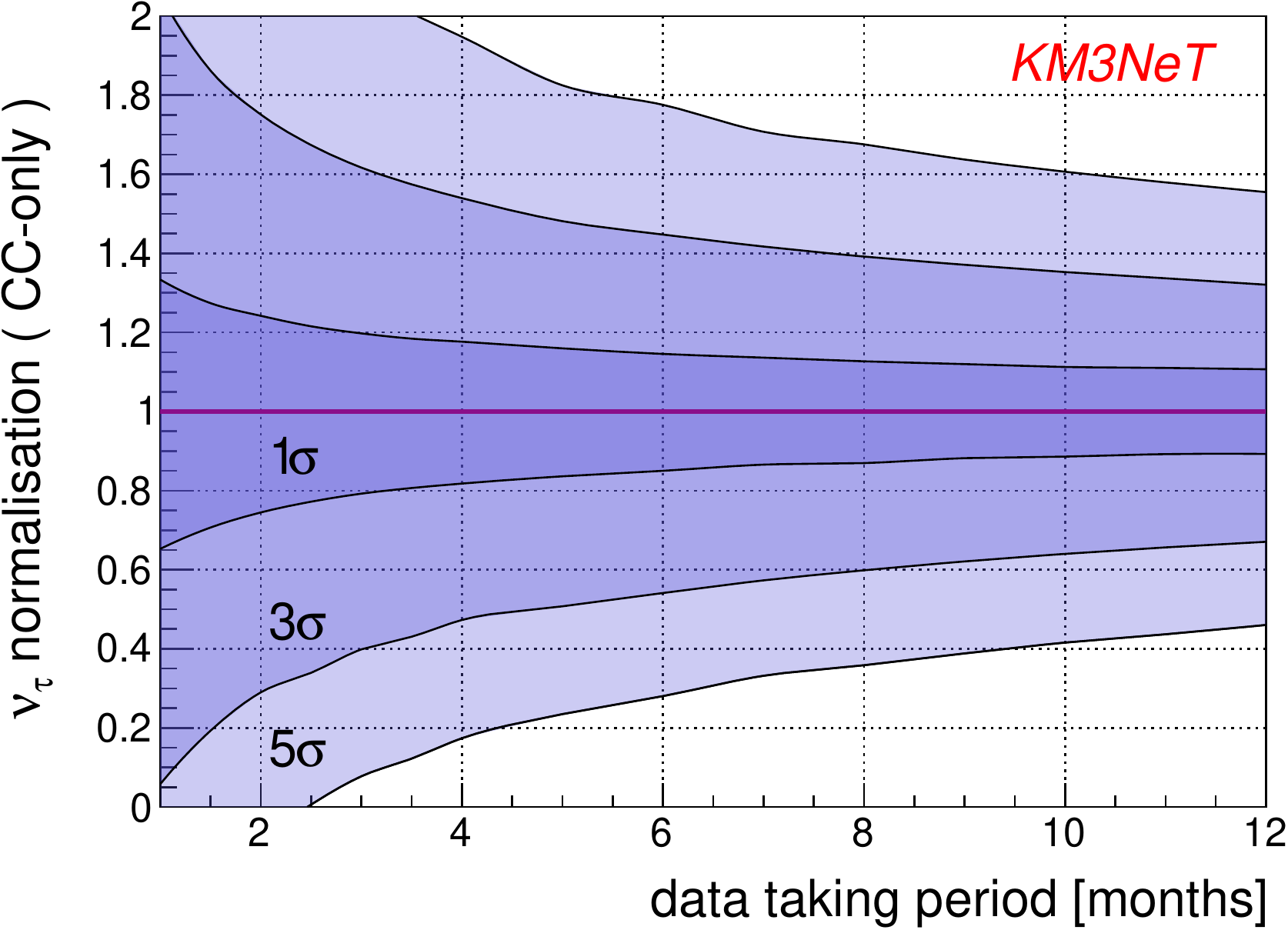}
		\label{fig:tauappearance2D}
	}	
	\caption{Sensitivity to $\nuantau$ appearance for CC and CC+NC normalisation scaling after one and three years of operation \protect\subref{fig:tauappearance1D}. Measurements from other experiments \cite{tau_appearance_opera_finalresults2018,tau_appearance_superk_2017,tau_appearance_icecube_2019} at 1$\sigma$ level are shown for comparison. In  \protect\subref{fig:tauappearance2D}, $\nuantau$ appearance sensitivity for CC scaling is presented as a function of data taking period.}
	\label{fig:tau_appearance}
\end{figure}

The sensitivity to $\nuantau$ appearance after one year and three years of operation for CC and CC+NC normalisation scaling is shown for a scan in $n_{\nuantau}$ in \autoref{fig:tauappearance1D}. In \autoref{fig:tauappearance2D}, the sensitivity for CC-only scaling is presented as a function of operation time. 

KM3NeT/ORCA will already be able to confirm the exclusion of non-appearance with high statistical significance with few months of data-taking. For CC the normalisation can be constrained to $\pm30\%$ at $3\sigma$-level and to $\pm 10\%$ at $1\sigma$-level after one year of data taking. After three years, the normalisation can be constrained to $\pm20\%$ at $3\sigma$-level, and to $\pm 7\%$ at $1\sigma$-level. The measured $\nuantau$ normalisation is robust against an incorrectly assumed sign of the still undetermined NMO. This enables KM3NeT/ORCA to measure $\nuantau$ appearance already during an early phase of construction \cite{icrc2019_ORCAearlymeasurements_strandberg_hallmann}.

\section{Conclusions}
\label{Sec:conclusion}
The importance of an independent study of neutrino oscillations, notably the determination of the NMO, has recently been reinforced as earlier hints, which favoured NO, are fading away in the light of latest combined results~\cite{Esteban:2020cvm,Kelly:2020fkv}.

The KM3NeT/ORCA sensitivity to atmospheric neutrino oscillation has been updated accounting for an optimised detector geometry and major improvements in neutrino trigger and reconstruction algorithms, and data analysis.
The trigger algorithm has been improved allowing to more efficiently collect neutrinos in the few-\si{GeV} energy range. The algorithms to select neutrino flavour-enriched samples have been optimised using multivariate analysis techniques. Finally, the models used in the statistical analysis have been refined with a realistic description of the systematic uncertainties.

The sensitivity to determine the NMO after three years of data taking was found to be \nmoNO\, (\nmoIO)\,$\sigma$ if the true NMO is NO (IO) and the other oscillation parameters are set to the current best estimates~\cite{globalFitEsteban}. The measurement precision on $\dmsq{32}$ and $\theta_{23}$ are \dmErrNO\ and \dThErrNO\ for NO, and \dmErrIO\ and \dThErrIO\ for IO. Finally, the unitary $3\times3$ neutrino mixing paradigm can be assessed by confronting the $\nuantau$ event rate to the expectation in this model. With three years of data taking, $\nuantau$ event rate variation larger than \SI{20}{\%} can be excluded at the 3$\sigma$ level.

\begin{acknowledgements}
The authors acknowledge the financial support of the funding agencies:
Agence Nationale de la Recherche (contract ANR-15-CE31-0020),
Centre National de la Recherche Scientifique (CNRS), 
Commission Europ\'eenne (FEDER fund and Marie Curie Program),
Institut Universitaire de France (IUF),
LabEx UnivEarthS (ANR-10-LABX-0023 and ANR-18-IDEX-0001),
Paris \^Ile-de-France Region,
France;
Shota Rustaveli National Science Foundation of Georgia (SRNSFG, FR-18-1268),
Georgia;
Deutsche Forschungsgemeinschaft (DFG),
Germany;
The General Secretariat of Research and Technology (GSRT),
Greece;
Istituto Nazionale di Fisica Nucleare (INFN),
Ministero dell'Universit\`a e della Ricerca (MIUR),
PRIN 2017 program (Grant NAT-NET 2017W4HA7S)
Italy;
Ministry of Higher Education Scientific Research and Professional Training,
ICTP through Grant AF-13,
Morocco;
Nederlandse organisatie voor Wetenschappelijk Onderzoek (NWO),
the Netherlands;
The National Science Centre, Poland (2015/18/E/ST2/00758);
National Authority for Scientific Research (ANCS),
Romania;
Ministerio de Ciencia, Innovaci\'{o}n, Investigaci\'{o}n y Universidades (MCIU): Programa Estatal de Generaci\'{o}n de Conocimiento (refs. PGC2018-096663-B-C41, -A-C42, -B-C43, -B-C44) (MCIU/FEDER), Severo Ochoa Centre of Excellence and MultiDark Consolider (MCIU), Junta de Andaluc\'{i}a (ref. SOMM17/6104/UGR), Generalitat Valenciana: Grisol\'{i}a (ref. GRISOLIA/2018/119) and GenT (ref. CIDEGENT/2018/034 and CIDEGENT/2019/043) programs, La Caixa Foundation (ref. LCF/BQ/IN17/11620019), EU: MSC program (ref. 713673),
Spain.

\end{acknowledgements}

\bibliographystyle{spphys}       
\bibliography{references.bib}

\begin{thebibliography}{10}
\providecommand{\url}[1]{{#1}}
\providecommand{\urlprefix}{URL }
\expandafter\ifx\csname urlstyle\endcsname\relax
  \providecommand{\doi}[1]{DOI \discretionary{}{}{}#1}\else
  \providecommand{\doi}{DOI \discretionary{}{}{}\begingroup
  \urlstyle{rm}\Url}\fi

\bibitem{MakiEtAl_1962}
B.~Maki, M.~Nakagawa, S.~Sakata, Prog. Theor. Phys. \textbf{28}, 870 (1962).
\newblock \doi{10.1143/PTP.28.870}

\bibitem{Pontecorvo_1968}
B.~Pontecorvo, Sov. Phys. JETP \textbf{26}, 984 (1968)

\bibitem{GribovEtAl_1969}
V.~Gribov, B.~Pontecorvo, Phys. Lett. B \textbf{28}, 493 (1969).
\newblock \doi{10.1016/0370-2693(69)90525-5}

\bibitem{PDG2018}
{M. Tanabashi et al. (Particle Data Group)}, Phys. Rev. D \textbf{98}(3),
  030001 (2018).
\newblock \doi{10.1103/PhysRevD.98.030001}

\bibitem{globalFitSalas}
P.F. de~Salas, D.V. Forero, C.A. Ternes, M.~T{\'o}rtola, J.W.F. Valle, Phys.
  Lett. \textbf{B782}, 633 (2018).
\newblock \doi{10.1016/j.physletb.2018.06.019}

\bibitem{globalFitEsteban}
I.~Esteban, M.C. Gonzalez-Garcia, A.~Hernandez-Cabezudo, M.~Maltoni,
  T.~Schwetz, Journal of High Energy Physics \textbf{01}, 106 (2019).
\newblock \doi{10.1007/JHEP01(2019)106}.
\newblock NuFIT 4.1 (2019), www.nu-fit.org

\bibitem{globalFitCapozzi}
F.~Capozzi, E.~Lisi, A.~Marrone, A.~Palazzo, Prog. Part. Nucl. Phys.
  \textbf{102}, 48 (2018).
\newblock \doi{10.1016/j.ppnp.2018.05.005}

\bibitem{Esteban:2020cvm}
I.~Esteban, M.C. Gonzalez-Garcia, M.~Maltoni, T.~Schwetz, A.~Zhou, JHEP
  \textbf{09}, 178 (2020).
\newblock \doi{10.1007/JHEP09(2020)178}

\bibitem{Kelly:2020fkv}
K.J. Kelly, P.A. Machado, S.J. Parke, Y.F. Perez~Gonzalez,
  R.~Zukanovich-Funchal, Phys. Rev. D \textbf{103}(1), 013004 (2021).
\newblock \doi{10.1103/PhysRevD.103.013004}

\bibitem{hep-ph_T2K_AbeEtAl_2020}
{K. Abe et al. (T2K Collaboration)}, Nature \textbf{580}(7803), 339 (2020).
\newblock \doi{10.1038/s41586-020-2177-0}

\bibitem{hep-ph_NOvA_AceroEtAl_2019}
{M.~A. Acero et al. (NOvA Collaboration)}, Phys. Rev. Lett. \textbf{123}(15),
  151803 (2019).
\newblock \doi{10.1103/PhysRevLett.123.151803}

\bibitem{aurisano_adam_2018_1286760}
A.~Aurisano.
\newblock {Recent Results from MINOS and MINOS+} (2018).
\newblock \doi{10.5281/zenodo.1286760}

\bibitem{hep-ph_SuperKamiokande_2018}
{K. Abe et al. (Super-Kamiokande Collaboration)}, Phys. Rev. D \textbf{97}(7),
  072001 (2018).
\newblock \doi{10.1103/PhysRevD.97.072001}

\bibitem{DeepCore2017}
{M.~G. Aartsen et al. (IceCube Collaboration)}, Phys. Rev. Lett.
  \textbf{120}(7), 071801 (2018).
\newblock \doi{10.1103/PhysRevLett.120.071801}

\bibitem{neutrino2020talk_T2K}
P.~Dunne.
\newblock {Latest Neutrino Oscillation Results from T2K} (2020).
\newblock \doi{10.5281/zenodo.4154355}

\bibitem{neutrino2020talk_NOvA}
A.~Himmel.
\newblock {New Oscillation Results from the NOvA Experiment} (2020).
\newblock \doi{10.5281/zenodo.3959581}

\bibitem{tau_appearance_opera_discovery2015}
{N. Agafonova et al. (OPERA Collaboration)}, Phys. Rev. Lett. \textbf{115}(12),
  121802 (2015).
\newblock \doi{10.1103/PhysRevLett.115.121802}

\bibitem{tau_appearance_opera_finalresults2018}
{N. Agafonova et al. (OPERA Collaboration)}, Phys. Rev. Lett. \textbf{120}(21),
  211801 (2018).
\newblock \doi{10.1103/PhysRevLett.120.211801}.
\newblock [Erratum: Phys.Rev.Lett. 121, 139901 (2018)]

\bibitem{tau_appearance_superk_2017}
{Z. Li et al. (Super-Kamiokande Collaboration)}, Phys. Rev. D \textbf{98}(5),
  052006 (2018).
\newblock \doi{10.1103/PhysRevD.98.052006}

\bibitem{tau_appearance_icecube_2019}
{M.~G. Aartsen et al. (IceCube Collaboration)}, Phys. Rev. D \textbf{99}(3),
  032007 (2019).
\newblock \doi{10.1103/PhysRevD.99.032007}

\bibitem{Akhmedov2013}
E.K. Akhmedov, S.~Razzaque, A.{\relax Yu}. Smirnov, Journal of High Energy
  Physics \textbf{02}, 82 (2013).
\newblock \doi{10.1007/JHEP02(2013)082}

\bibitem{Wolfenstein1978}
L.~Wolfenstein, Phys. Rev. D \textbf{17}, 2369 (1978).
\newblock \doi{10.1103/PhysRevD.17.2369}

\bibitem{Mikheev:1986gs}
S.P. Mikheyev, A.Y. Smirnov, Sov. J. Nucl. Phys. \textbf{42}, 913 (1985)

\bibitem{LoI}
{S. Adri{\'a}n-Mart{\'i}nez et al. (KM3NeT Collaboration)}, Journal of Physics
  G \textbf{43}(8) (2016).
\newblock \doi{10.1088/0954-3899/43/8/084001}

\bibitem{Aiello2018}
{S. Aiello et al. (KM3NeT Collaboration)}, JINST \textbf{13}(05), P05035
  (2018).
\newblock \doi{10.1088/1748-0221/13/05/P05035}

\bibitem{Aiello2020KM3NeTdeployment}
{S. Aiello et al. (KM3NeT Collaboration)}, JINST \textbf{15}(11), P11027
  (2020).
\newblock \doi{10.1088/1748-0221/15/11/P11027}

\bibitem{gSeaGen_paper}
{S. Aiello et al. (KM3NeT Collaboration)}, Comput. Phys. Commun. \textbf{256},
  107477 (2020).
\newblock \doi{10.1016/j.cpc.2020.107477}

\bibitem{GENIE}
C.~Andreopoulos, et~al., Nucl. Instrum. Meth. A \textbf{614}, 87 (2010).
\newblock \doi{10.1016/j.nima.2009.12.009}

\bibitem{GENIE_manual}
C.~Andreopoulos, et~al.
\newblock {The GENIE Neutrino Monte Carlo Generator: Physics and User Manual}
  (2015).
\newblock {arXiv:1510.05494 (hep-ph)}

\bibitem{Honda2015}
M.~Honda, M.S. Athar, T.~Kajita, K.~Kasahara, S.~Midorikawa, Phys. Rev. D
  \textbf{92}, 023004 (2015).
\newblock \doi{10.1103/PhysRevD.92.023004}

\bibitem{KM3Sim}
A.G. Tsirigotis, A.~Leisos, S.E. Tzamarias, Nucl. Instrum. Meth. A
  \textbf{626-627}, S185 (2011).
\newblock \doi{10.1016/j.nima.2010.06.258}

\bibitem{MUPAGE}
G.~Carminati, A.~Margiotta, M.~Spurio, Comput. Phys. Commun. \textbf{179}, 915
  (2008).
\newblock \doi{10.1016/j.cpc.2008.07.014}

\bibitem{Bailey:2002uj}
D.~Bailey, {Monte Carlo tools and analysis methods for understanding the
  ANTARES experiment and predicting its sensitivity to dark matter}.
\newblock Ph.D. thesis, University of Oxford (2002)

\bibitem{AntaresSimulation2020}
{A. Albert et al. (ANTARES Collaboration)}, JCAP \textbf{01}, 064 (2021).
\newblock \doi{10.1088/1475-7516/2021/01/064}

\bibitem{depthIntensityRelationPaper}
{M. Ageron et al. (KM3NeT Collaboration)}, Eur. Phys. J. C \textbf{80}(2), 99
  (2020).
\newblock \doi{10.1140/epjc/s10052-020-7629-z}

\bibitem{SteffenThesis}
S.~Hallmann, {Sensitivity to atmospheric tau-neutrino appearance and
  all-flavour search for neutrinos from the Fermi Bubbles with the deep-sea
  telescopes KM3NeT/ORCA and ANTARES}.
\newblock Ph.D. thesis, Friedrich-Alexander-Universit{\"a}t
  Erlangen-N{\"u}rnberg (FAU) (2021).
\newblock
  \urlprefix\url{https://nbn-resolving.org/urn:nbn:de:bvb:29-opus4-157495}

\bibitem{LiamThesis}
L.~Quinn, {Neutrino Mass Hierarchy Determination with KM3Net/ORCA}.
\newblock Ph.D. thesis, Aix-Marseille University (2018).
\newblock \urlprefix\url{http://hal.in2p3.fr/tel-02265297}

\bibitem{JannikThesis}
J.~Hofestädt, {Measuring the neutrino mass hierarchy with the future
  KM3NeT/ORCA detector}.
\newblock Ph.D. thesis, Friedrich-Alexander-Universit{\"a}t
  Erlangen-N{\"u}rnberg (FAU) (2017).
\newblock
  \urlprefix\url{https://nbn-resolving.org/urn:nbn:de:bvb:29-opus4-82770}

\bibitem{nutau_anutau_xsection}
Y.S. Jeong, M.H. Reno, Phys. Rev. D \textbf{82}, 033010 (2010).
\newblock \doi{10.1103/PhysRevD.82.033010}

\bibitem{IntrinsicPaper}
{S. Adri\'an-Mart\'\i{}nez et al. (KM3NeT Collaboration)}, JHEP \textbf{05},
  008 (2017).
\newblock \doi{10.1007/JHEP05(2017)008}

\bibitem{Breiman2001}
L.~Breiman, Machine Learning \textbf{45}(1), 5 (2001).
\newblock \doi{10.1023/A:1010933404324}

\bibitem{Zeller2012}
J.A. Formaggio, G.P. Zeller, Rev. Mod. Phys. \textbf{84}, 1307 (2012).
\newblock \doi{10.1103/RevModPhys.84.1307}

\bibitem{OscProb}
J.~Coelho.
\newblock {OscProb Neutrino Oscillation Calculator}.
\newblock \urlprefix\url{https://github.com/joaoabcoelho/OscProb}

\bibitem{CowanEtAl_2011}
G.~Cowan, K.~Cranmer, E.~Gross, O.~Vitells, Eur. Phys. J. C \textbf{71}, 1554
  (2011).
\newblock \doi{10.1140/epjc/s10052-011-1554-0}.
\newblock [Erratum: Eur.Phys.J.C 73, 2501 (2013)]

\bibitem{hep-ph_BakerEtAl_1984}
S.~Baker, R.D. Cousins, Nucl. Instrum. Meth. \textbf{221}, 437 (1984).
\newblock \doi{10.1016/0167-5087(84)90016-4}

\bibitem{BarrEtAl_2006}
G.D. Barr, T.K. Gaisser, S.~Robbins, T.~Stanev, Phys. Rev. D \textbf{74},
  094009 (2006).
\newblock \doi{10.1103/PhysRevD.74.094009}

\bibitem{ANTARES_AlbertEtAl_2018}
{A. Albert et al. (ANTARES Collaboration)}, Eur. Phys. J. C \textbf{78}(8), 669
  (2018).
\newblock \doi{10.1140/epjc/s10052-018-6132-2}

\bibitem{premModel}
A.M. Dziewonski, D.L. Anderson, Phys. Earth Planet. Interiors \textbf{25}, 297
  (1981).
\newblock \doi{10.1016/0031-9201(81)90046-7}

\bibitem{icrc2019_ORCAearlymeasurements_strandberg_hallmann}
B.~Strandberg, S.~Hallmann, PoS \textbf{ICRC2019}, 1019 (2019).
\newblock \doi{10.22323/1.358.1019}

\end{thebibliography}
\end{document}